# INTELLIGENT AUTOMATION FOR FDI FACILITATION: Optimizing Tariff Exemption Processes with OCR And Large Language Models

MUHAMMAD SUKRI BIN RAMLI
Asia School of Business
Kuala Lumpur, Malaysia
Email: m.binramli@sloan.mit.edu

**Abstract**

Tariff exemptions are fundamental to attracting Foreign Direct Investment (FDI) into the manufacturing sector, though the associated administrative processes present areas for optimization for both investing entities and the national tax authority. This paper proposes a conceptual framework to empower tax administration by leveraging a synergistic integration of Optical Character Recognition (OCR) and Large Language Model (LLM) technologies. The proposed system is designed to first utilize OCR for intelligent digitization, precisely extracting data from diverse application documents and key regulatory texts such as tariff orders. Subsequently, the LLM would enhance the capabilities of administrative officers by automating the critical and time-intensive task of verifying submitted HS Tariff Codes for machinery, equipment, and raw materials against official exemption lists. By enhancing the speed and precision of these initial assessments, this AI-driven approach systematically reduces potential for non-alignment and non-optimized exemption utilization, thereby streamlining the investment journey for FDI companies. For the national administration, the benefits include a significant boost in operational capacity, reduced administrative load, and a strengthened control environment, ultimately improving the ease of doing business and solidifying the nation's appeal as a premier destination for high-value manufacturing FDI.

.
1. Introduction

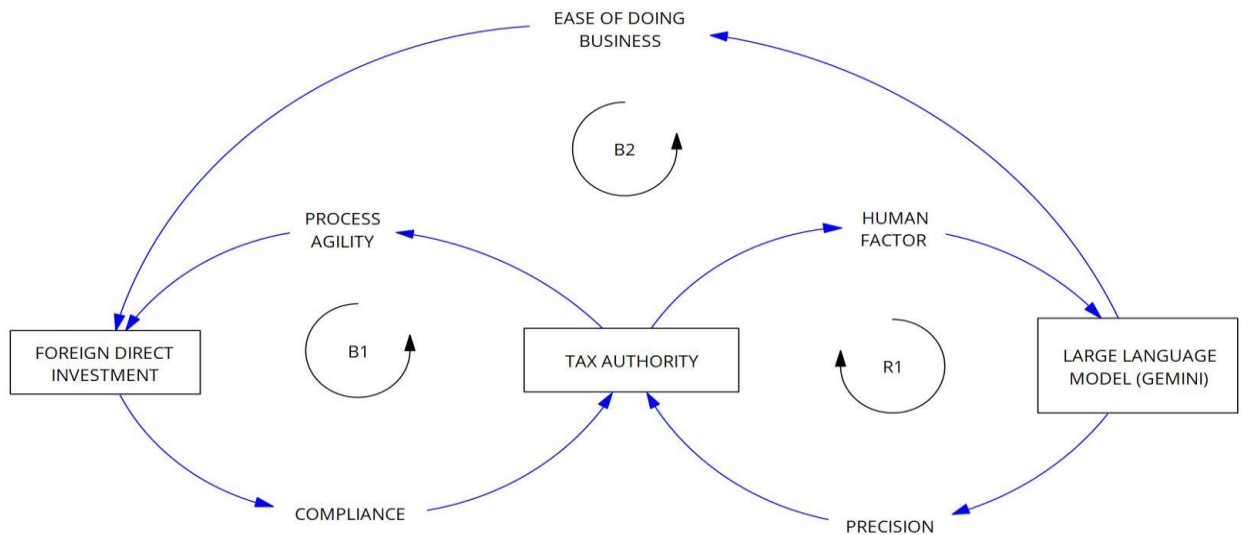

Figure 1: Causal Loop Diagram Illustrating relationship between FDI, Tax Authority and LLM.

The causal loop diagram on figure 1 illustrates how integrating a Large Language Model like Gemini is designed to cultivate a predominantly positive feedback system, ultimately enhancing Foreign Direct Investment (FDI). The interplay of key reinforcing and balancing loops R1, B1, and B2 describes the mechanics of this transformation. The primary driver of enhancement is the reinforcing loop R1. This loop begins with Gemini significantly boosting the "Precision" of the "Tax Authority," for instance, in verifying tariff exemptions. Simultaneously, Gemini positively influences the "Human Factor" by reducing administrative load and ensuring consistent application of rules. These enhancements strengthen the overall capacity and effectiveness of the "Tax Authority". This improved authority, in turn, leads to greater reliance on and further optimization of Gemini, creating a virtuous cycle that continuously reinforces "Precision" and human effectiveness. The direct outcomes are increased "Process Agility" in handling tax-related applications and improved "Compliance" with regulations. As the "Tax Authority" becomes more efficient and the investment environment more predictable (due to R1), "Foreign Direct Investment" is encouraged. However, this growth introduces dynamic factors, represented by the balancing loop B1. Rising FDI can place increased demands on the tax system, potentially influencing "Process Agility" and the mechanisms ensuring "Compliance". B1 highlights this aspect: for the system to maintain its attractiveness, the enhancements driven by R1 must effectively absorb these pressures, ensuring the "Tax Authority" can scale its improved capabilities. The combined effect of enhanced "Process

Agility" and "Compliance," alongside a more effective "Human Factor," directly contributes to an improved "Ease of Doing Business" (EoDB), a critical factor for attracting FDI. The balancing loop B2, shown directly on "Ease of Doing Business," introduces a longer-term, nuanced dynamic. It suggests that as Gemini's integration successfully optimizes processes (like the tariff exemption task) and contributes to a highly satisfactory level of EoDB, the specific strategic demands on Gemini for that particular task might achieve a stable optimization or see diminishing marginal returns from further intensive refinement. Figure 2, titled "Comparison of HS Code Checking Speed: Manual vs. LLM-Assisted," hypothetically presents a bar chart illustrating the disparity in processing time for 300 Harmonized System (HS) codes between manual verification and an LLM-assisted approach. The "Manual Officer Check" bar indicates a total processing duration of 450.0 minutes. Conversely, the "LLM-Assisted Check" bar demonstrates a substantially reduced processing time of 5.0 minutes for the identical task. The accompanying note clarifies that these data points are derived from hypothetical user-provided figures, with the manual process estimated at 1 minute 30 seconds per code and the LLM-assisted process at 5 minutes for all 300 codes, thereby underscoring the projected efficiency gains attainable through an LLM system in the context of Foreign Direct Investment (FDI) tariff exemption processing.

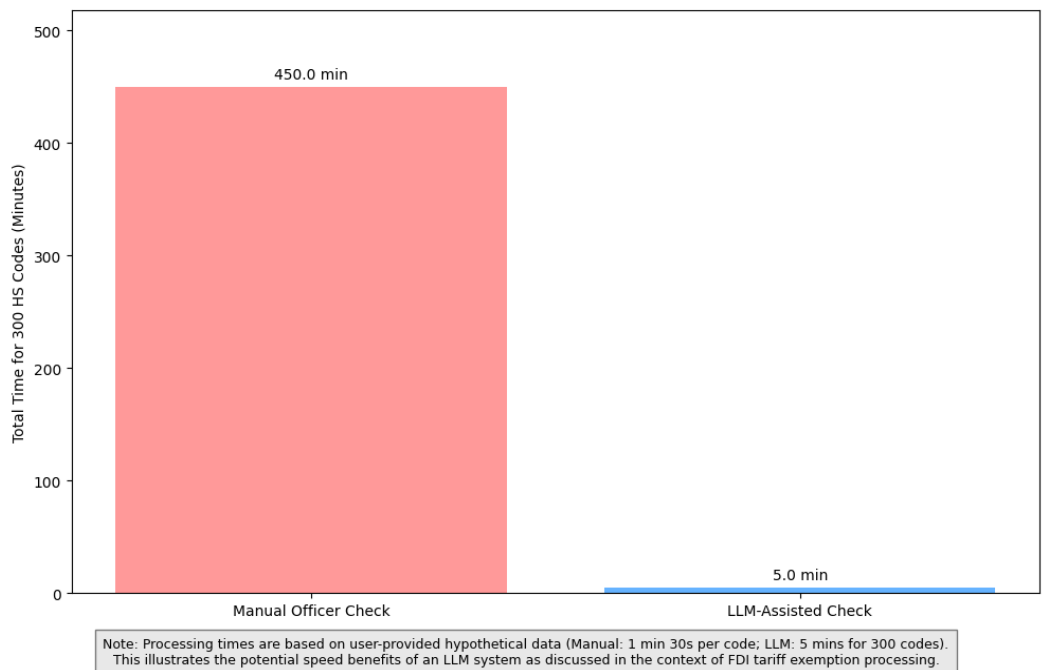

Figure 2: Speed comparison between manual vs LLM-Assisted HS Code Checking

### 1.1. Background: The Imperative of Manufacturing FDI for Malaysia's Economy

Malaysia's economic advancement relies on its national development blueprints, notably the New Industrial Master Plan 2030 (NIMP 2030). This plan, as emphasized by the Ministry of Investment, Trade and Industry (2023), strategically leverages Foreign Direct Investment (FDI) to achieve its ambitious targets, a point reinforced by studies from Abdullah and Zolkornain (2021) and Athukorala (2019) highlighting FDI's critical role in the nation's economic growth. Within this framework, tariff exemptions act as a key fiscal incentive to attract and retain manufacturing FDI, thereby boosting industrial activity and technological progress, with the effective administration of such incentives being paramount, as discussed in broader tax administration analyses (Bird, 2015; Keen & Slemrod, 2017). Simultaneously, the nation is bolstering its technological infrastructure by investing in advanced Artificial Intelligence capabilities, including access to potent Large Language Models like Google's Gemini via platforms such as Google Suites. This dedication to AI adoption, as outlined in national digital strategies, is fundamental to modernizing the public sector and improving service delivery, reflecting global trends in public service AI adoption. Demonstrating this commitment, an "AI at Work" pilot program involved 270 public officers in December 2024, which was significantly expanded by February 2025 under the "AI at Work 2.0" initiative to potentially include up to 445,000 officers.

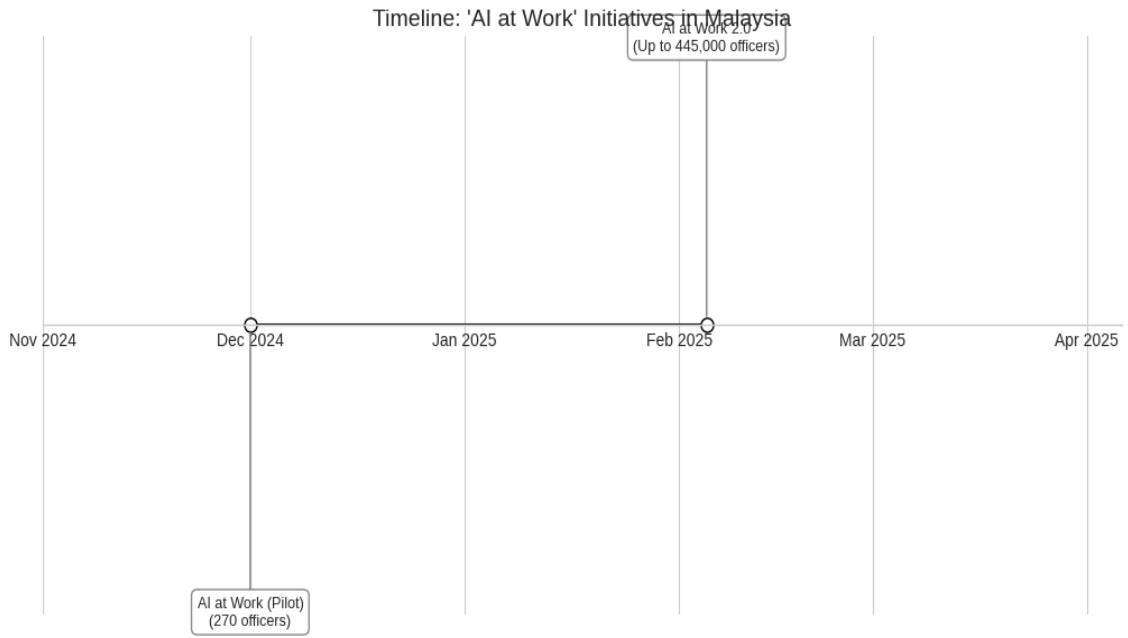

Figure 3: Timeline 'AI at Work' initiative in Malaysia

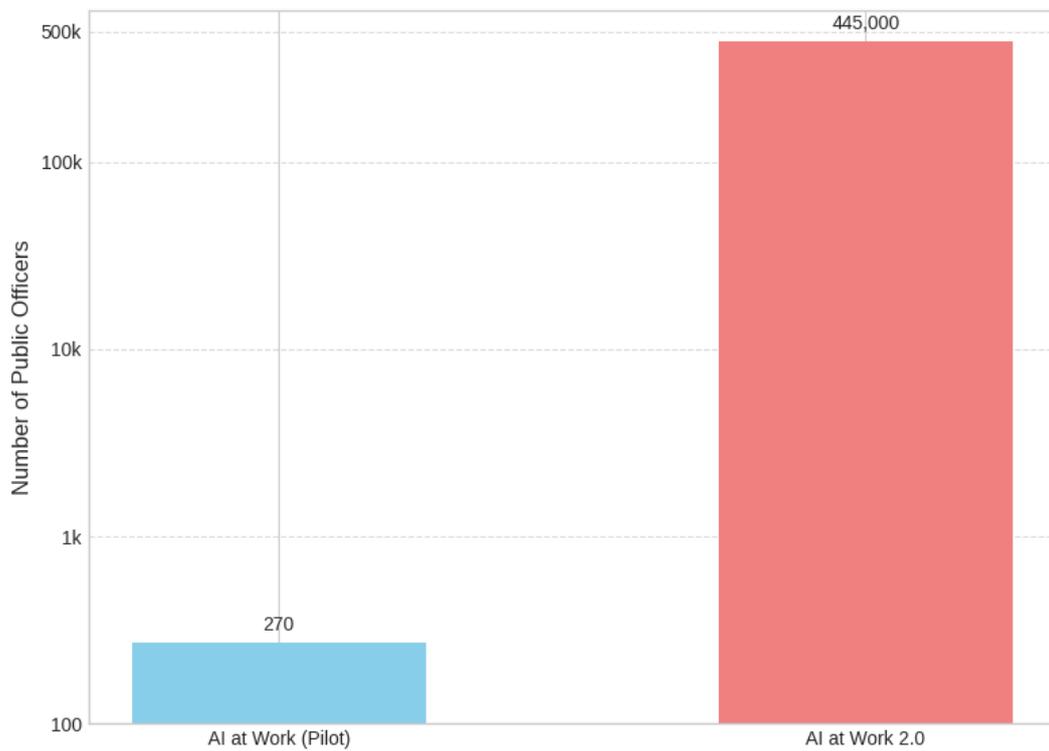

Figure 4: Number of Public Officers participate for 'AI at Work'

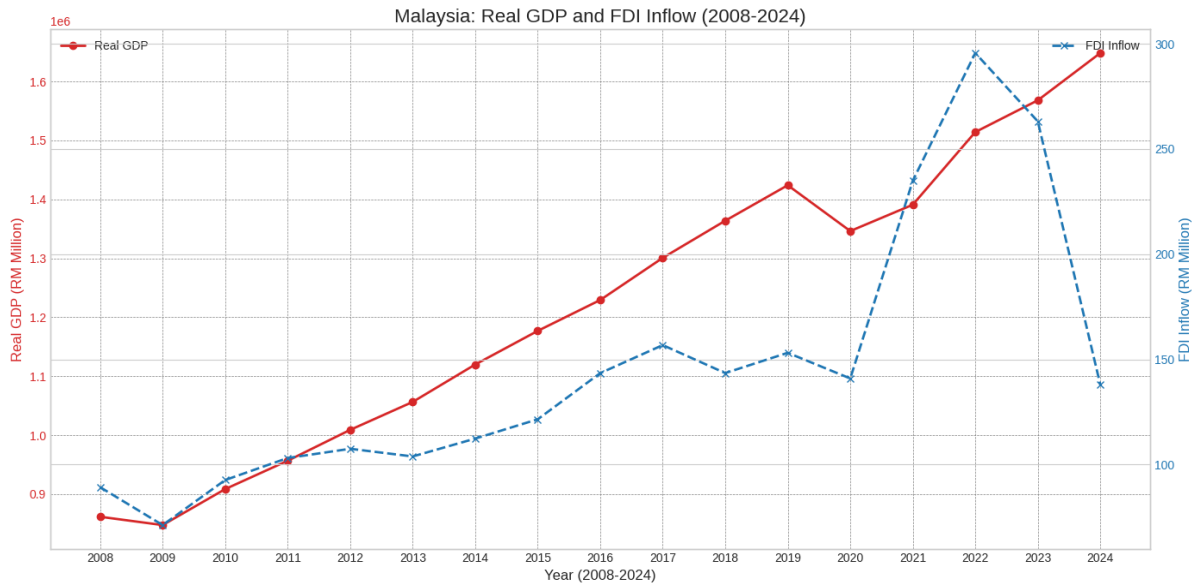

Figure 5: Relationship between Malaysia Real GDP and FDI Inflow (2008-2024)

From 2008 to 2024, Malaysia's economic landscape, as depicted by its Real GDP and Foreign Direct Investment (FDI) inflow, showcases two distinct trends sourced from the nation's Official Open Data Portal (data.gov.my). Real GDP consistently demonstrated a robust upward trajectory, reflecting sustained long-term economic growth over the seventeen-year period, effectively doubling its size by 2024. While temporary dips occurred due to global economic events in 2009 and 2020, GDP swiftly recovered and continued its expansion, albeit with a brief moderation in growth around 2020. In stark contrast, FDI inflow exhibited considerable short-term volatility, characterized by significant fluctuations, including pronounced peaks in 2016-2017 and a dramatic surge in 2022. However, this surge was followed by a sharp contraction by 2024, indicating a less stable and predictable pattern compared to GDP. The most recent years highlight a notable divergence, where Real GDP continued its strong growth, reaching new highs, even as FDI inflows experienced a significant decrease from their exceptional 2022 peak, suggesting a potential decoupling of these two indicators in Malaysia's recent economic performance.

### 1.2. The Challenge: Navigating Tariff Exemption Complexity

Recognizing the clear benefits of tariff exemptions towards the volume of FDI, Malaysian tax administration bodies, particularly the Royal Malaysian Customs Department (RMCD) and the Malaysian Investment Development Authority (MIDA), are dedicated to continuously improving the efficiency and accuracy of processing the substantial volume of applications for these exemptions. The complexity inherent in tariff classification, particularly ensuring Harmonized System (HS) codes align with national directives as detailed in various Customs Orders and Regulations, presents a persistent challenge, a complexity noted in academic discussions on tariff classification uncertainty and its trade impact (Felbermayr, Teti, & Yalcin, 2021; Waugh, 1997). These challenges in managing complex regulatory information are not uncommon in public administration globally. However, the nation's existing strategic investment in AI, specifically the advanced capabilities of Google's Gemini LLM, presents a significant opportunity. This technological infrastructure, as discussed in contexts of automation and digital transformation (Acemoglu & Restrepo, 2019; Brynjolfsson & McAfee, 2017), can be leveraged to address these administrative complexities, offering pathways to automate and augment the decision-making processes involved in tariff exemptions.

### 1.3. Research Aim and Objectives

The central aim of this research is to investigate and propose a conceptual framework for an Optical Character Recognition (OCR)-powered system, thoughtfully integrated with Google's Gemini Large Language Model (LLM). This framework is primarily envisioned to enhance the capabilities of the national tax authority in expediting, securing, and optimizing tariff exemption processes pertinent to manufacturing Foreign Direct Investment (FDI). This proposal, leveraging LLM and OCR for FDI, is considered exceptionally high-impact due to its direct relevance to substantial investments. The technological core of this proposed framework lies in combining the historical development and modern applications of OCR technology for intelligent document processing with the advanced reasoning and language understanding of Gemini. This aligns seamlessly with broader public sector modernization efforts and the overarching goal of creating significant public value through smart technologies. To achieve this aim, the research pursues several key objectives. Firstly, it details current administrative workflows and identify specific areas for optimization encountered in verifying FDI tariff exemptions, with a particular focus on the nuances of HS Code classification against official national lists and regulations. Secondly, the study pinpoints specific operational considerations within these workflows where the advanced language understanding and reasoning capabilities of Gemini, synergistically combined with OCR for document processing, can directly assist officers in their verification tasks. This may involve adapting language models to the specific regulatory domain. Thirdly, the research explores the specific functionalities of both OCR and Google's Gemini LLM, such as its multimodality and contextual understanding, that are most relevant to these complex verification tasks. Fourthly, a conceptual framework for this AI-driven system be proposed, tailored for effective use by these administrative bodies. Finally, the research analyzes how such a system can improve administrative efficiency, enhance accuracy in decision-making while considering the importance of interpretable models in high-stakes decisions, strengthen control over potential exemption non-alignment, and consequently, reduce non-compliance factors for FDI companies, thereby bolstering the overall investment climate and ensuring the system remains trustworthy and human-centric.

### 1.4. Research Questions

This study is guided by key research questions designed to explore the potential of administrative empowerment through AI. Primarily, it seeks to determine how Optical Character Recognition (OCR) technology and Google's Gemini Large Language Model can be specifically and effectively integrated into the existing workflow of the national tax administration. The objective is to understand how this integration can enhance the accuracy and efficiency of verifying HS Tariff Codes for FDI exemption applications against official Malaysian directives and regulatory texts, a field informed by surveys on deep learning for legal analytics and LLMs in RegTech (Klanrit, Lee, & Kim, 2023). This involves considering models for human-AI interaction and guidelines for designing such collaborative systems (Parasuraman, Sheridan, & Wickens, 2000; Amershi et al., 2019). Furthermore, the research investigates the key features of a Gemini-powered system that would directly contribute to reducing officer fatigue and improving the consistency of decision-making in the complex process of granting tariff exemptions. This involves understanding how AI can support human decision-making as explored by Benbasat and Lim (2000) and Klein (2008), maintain officer situation awareness as theorized by Endsley (2017), and provide transparent, explainable outputs to ensure trust and consistency, drawing on insights from surveys on explainable AI and its role in decision-making (Guidotti et al., 2018; Wang et al., 2021).

### 1.5. Scope and Delimitations

The scope of this research is intentionally focused to ensure depth and relevance. The primary area of investigation concerns tariff exemptions specifically available to manufacturing Foreign Direct Investment (FDI) within the national context, a sector highlighted as critical in national economic plans. Technologically, the study concentrates on the application of Optical Character Recognition (OCR) for digitizing relevant documentation and the capabilities of Google's Gemini Large Language Model. These technologies be examined primarily from the perspective of their use by, or direct support for, the national tax administration bodies, aligning with national strategies for digital transformation and AI adoption in the public sector. The research does not extend to a full technical implementation or the development of new algorithms but focus on the conceptual framework, operational integration strategies, and the anticipated administrative impact.

### 1.6. Significance of the Study

The significance of this study is multifaceted, particularly for national tax administration. It proposes a tangible pathway towards modernization by leveraging a government-acquired, advanced AI tool Google's Gemini LLM. Such an initiative promises to enhance administrative efficiency, improve the accuracy of complex tariff exemption decisions, and provide greater control over the allocation of fiscal incentives. Ultimately, the successful integration of such a system is anticipated to contribute to the creation of demonstrable public value, a concept central to modern public management theories (Moore, 1995; Bryson, Crosby, & Bloomberg, 2014; Hartley, 2013), and aligns with best practices in digital government transformation that aim to enhance citizen services and overall governance (Twizeyimana & Andersson, 2019). By optimizing a critical component of the FDI facilitation process, the research also holds implications for strengthening the nation's attractiveness as an investment destination, fostering a more transparent and predictable regulatory environment for businesses, thereby supporting national economic goals.

## 2. Literature review

### 2.1. FDI Incentives and Manufacturing Growth in Malaysia

The economic development trajectory of Malaysia has historically been intertwined with strategic efforts to attract Foreign Direct Investment (FDI), particularly into its manufacturing sector. National blueprints, including the New Industrial Master Plan 2030 (Ministry of Investment, Trade and Industry, 2023), continue to emphasize FDI as a catalyst for industrial upgrading, export growth, and job creation. Academic research, such as studies by Abdullah and Zolkornain (2021) and Athukorala (2019), provides empirical evidence on the positive impact of FDI on Malaysia's economic growth and development. In 2024, Malaysia's position in the IMD World Competitiveness Ranking is a key indicator of its attractiveness as an investment destination within ASEAN. While Malaysia ranked 34th overall among 67 countries in the IMD 2024 report, it stood behind Singapore (1st), Thailand (25th), and Indonesia (27th) among its Southeast Asian peers. The efficacy of fiscal incentives, prominently featuring tariff exemptions, plays a significant role in Malaysia's strategy to enhance its competitiveness as an investment destination (Malaysian Investment Development Authority, various years). The effectiveness of these incentives is crucial, as reflected in broader discussions on taxation within the ASEAN region and its impact on business decisions (Sawyer, 2018).

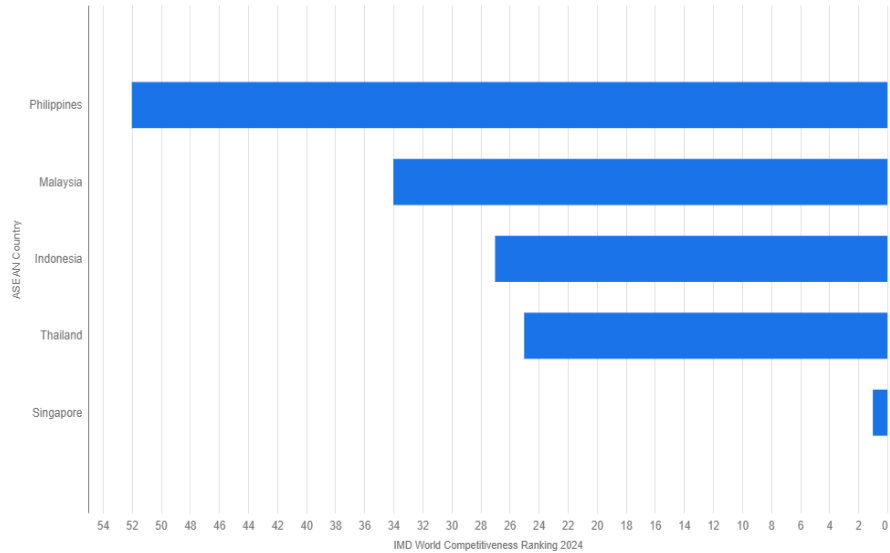

Figure 6: Malaysia Ranking for IMD World Competitiveness 2024 among ASEAN Countries

### 2.2. The Critical Role of Tax Administration in FDI Facilitation

| Metric | Effective Tax Administration | Ineffective Tax Administration |
| --- | --- | --- |
| FDI Attraction | High (Attracts & Retains Capital) | Low (Deters Investment) |
| Investor Confidence | High (Predictable, Transparent) | Low (Uncertainty, High Costs) |
| Operational Costs | Low (Smooth processing, incentives) | High (Delays, classification issues) |
| Economic Growth | Positive (Job creation, development) | Negative (Missed opportunities) |

Figure 7: Tax administration effectiveness metrics

Effective tax administration is a cornerstone of a favorable investment climate and plays a critical role in the successful facilitation of Foreign Direct Investment. International experience, as reviewed by scholars like Bird (2015) and Keen and Slemrod (2017) on optimal tax administration, demonstrates that efficient, transparent, and predictable tax systems are paramount for attracting and retaining foreign capital. For the tax authority, the processing of FDI-related incentives, such as tariff exemptions, such as tariff exemptions as outlined in their Customs Orders and Regulations, directly impacts the operational realities and financial viability of investing companies. Complexities and uncertainties in areas like tariff classification can introduce significant delays and costs, thereby potentially undermining the intended benefits of fiscal incentives, a point underscored by research on tariff-classification uncertainty and its effects on international trade (Felbermayr, Teti, & Yalcin, 2021). Modernization efforts within customs, often highlighted by bodies like the World Customs Organization and discussed in the context of AI's role in trade facilitation (Corteva, 2023), are therefore essential for ensuring that tax administration acts as an enabler, rather than an impediment, to FDI.

### 2.3. Manual Processes in Public Administration: Bottlenecks and Risks

Many public administration systems globally continue to manage processes with inherent manual aspects, which can present areas for optimization and introduce dynamic factors. The presence of such processes in critical functions, like the assessment of applications for fiscal incentives, can influence processing times, accuracy rates, and transparency (Bostrom & Heinen, 1977). The considerations associated with implementing and managing eGovernment solutions often stem from the need to evolve these established manual systems and manage factors highlighted in government IT project assessments. Optical Character Recognition (OCR) technology plays a pivotal role in bridging the gap between these traditional paper-based methods and the push towards digitalization. By converting physical documents into editable and searchable digital data, OCR helps eliminate the need for manual data entry, significantly reducing processing times and minimizing human error. While workplace automation offers a pathway to mitigate these issues by taking over repetitive task, the failure to effectively integrate technology and redesign work processes can perpetuate existing problems or even introduce new ones. Effective socio-technical design, as explored by Mumford (2006), is crucial to ensure that the move away from manual systems genuinely enhances efficiency and reduces risks such as inconsistent decision-making or susceptibility to misuse.

### 2.4. OCR Technology: From Document Digitization to Intelligent Data Capture

Optical Character Recognition (OCR) technology has evolved significantly from its origins as a tool for basic document digitization to a sophisticated component of intelligent data capture systems. The historical development of OCR, as chronicled by researchers like Mori, Suen, and Yamamoto (1992), laid the groundwork for modern engines such as Tesseract, an open-source option whose overview was provided by Smith (2007), which are capable of converting various document types into machine-readable text with increasing accuracy. More recently, the field has advanced towards "Intelligent Document Processing" (IDP), where OCR is combined with AI and machine learning techniques to not only extract text but also to understand its context, classify information, and validate data, as surveyed by Souza, Batista, and Silva (2021). This capability is crucial for automating the initial stages of information processing in complex administrative workflows, transforming unstructured or semi-structured data from scanned documents such as application forms, supporting invoices, and technical specifications into structured data amenable to further digital analysis and processing. OCR thus serves as a key enabler for subsequent automation and the reinstatement of labor to new tasks (Acemoglu & Restrepo, 2019).

### 2.5. Large Language Models (LLMs) in Regulatory Technology (RegTech): Focus on Google's Gemini

Large Language Models (LLMs) have emerged as transformative technologies with significant potential in Regulatory Technology (RegTech). These models, built upon architectures like the Transformer (Wolf et al., 2020) and exemplified by foundational models which showed few-shot learning capabilities (Brown et al., 2020) and more recent advancements like Google's Gemini, exhibit remarkable capabilities in understanding, generating, and reasoning with human language, as discussed in works on the opportunities and risks of such models (Bommasani et al., 2021) and comprehensive surveys of LLMs (Zhao et al., 2023).

As of early 2025, Google's Gemini is recognized for its advanced reasoning and contextual understanding strengths, details of which are regularly updated via Google AI and Google Research publications. Crucially for the application discussed herein, Gemini is selected not only for these general capabilities but specifically for its robust multimodal understanding. This is particularly advantageous for the Malaysian customs context, especially for Harmonized System (HS) Code recognition and classification, where application dockets for tariff exemptions often comprise a mix of typed text, scanned official forms that may include seals or stamps, and potentially complex tables or diagrams within technical specifications. Gemini's capacity, as detailed in contemporary reports from Google AI and Google Research, to holistically process and interpret such mixed-media documents offers a more comprehensive analytical starting point than LLMs primarily optimized for plain text, thereby enhancing the accuracy of data extraction and subsequent verification by systems like the proposed Gemini-Powered Verification Assistant (GPVA). Furthermore, if official Google sources by early 2025 indicate superior performance in handling multilingual content relevant to Malaysia (such as Bahasa Malaysia, English, and other scripts encountered in trade documentation) or specific advancements in few-shot learning for regulatory interpretation, these would be significant additional justifications for its selection, directly contributing to more precise HS code identification.

The potential deployment of Gemini via Google Suites and Google Cloud infrastructure provides a scalable platform for public sector applications, including in jurisdictions like Malaysia which may leverage such cloud services for public sector solutions and data residency. The capacity of these LLMs to be adapted or fine-tuned for specific domains, a topic explored in surveys on specialized LLMs (Liao et al., 2023), further enhances their utility, making them powerful tools for tasks previously considered too complex for automation and impacting specialized labor markets (Eloundou et al., 2023).

The application of advanced LLMs like Gemini extends naturally to compliance checking, the interpretation of complex regulatory texts, and the provision of sophisticated support for expert decision-making. Surveys in the RegTech domain (Zheng, Zhang, & Chen, 2021) and legal AI (Klanrit, Lee, & Kim, 2023; Zhong et al., 2020) highlight how LLMs can parse intricate legal and regulatory documents, identify relevant clauses, and assist in ensuring adherence to multifaceted compliance requirements. For instance, domain-specific models like LEGAL-BERT (Chalkidis et al., 2020) demonstrate the efficacy of fine-tuning LLMs for specialized legal language. In the context of customs tariff classification, AI's role in legal interpretation is already being explored (Fedi, 2018). Specifically, LLMs like Gemini can process vast amounts of customs regulations and official guidelines, enabling them to compare identified HS Codes against pre-defined lists of eligible versus non-eligible codes for specific tariff exemptions. This capability allows the system to flag potential discrepancies or confirm eligibility, significantly easing the task of customs officers by providing automated pre-screening, enhancing classification accuracy, and ensuring consistency across assessments. LLMs can serve as intelligent decision support systems, augmenting human capabilities in line with research trends in this area (Eom, 2009; Benbasat & Lim, 2000), helping officers navigate vast amounts of regulatory information and dedicate more focus to complex or ambiguous cases requiring human judgment. However, the challenge of capturing the "tacit dimension" of expert knowledge, as described by Polanyi (1966), remains a consideration for LLM application. To be effective and trustworthy in such roles, particularly in high-stakes public administration decisions, the explainability of these models is crucial, a field actively being researched through surveys on XAI methods (Adadi & Berrada, 2018; Guidotti et al., 2018).

### 2.6. AI Adoption in Global Customs and Tax Modernization: Trends and Case Studies

The adoption of Artificial Intelligence is an accelerating trend in the modernization efforts of customs and tax administrations worldwide, as these organizations seek to enhance efficiency, improve risk management, and facilitate legitimate trade and revenue collection. AI applications in public services are becoming increasingly common, with typologies of such applications being developed (Barcevičius, Cibaitė, & Codagnone, 2019), and international bodies like the OECD (2019) examining the societal and governmental implications of AI. While comprehensive, publicly documented case studies of advanced LLMs like Google's Gemini being fully deployed in customs operations for tasks such as nuanced tariff exemption processing were still emerging as of early 2025, the trajectory is clear. Existing literature points to AI, more broadly, as a game changer for trade facilitation and compliance in customs (Corteva, 2023). Malaysia is actively pursuing AI integration across its public sector, including revenue and border agencies. This commitment is underscored by Malaysia's National Artificial Intelligence Roadmap 2021-2025 (AI-Rmap) and the recent establishment of the National AI Office (NAIO) in December 2024. These initiatives aim to accelerate AI adoption, foster innovation, and enable more sophisticated AI deployments in government services, building upon earlier explorations of AI for citizen services. The World Customs Organization (WCO News, various years) often features discussions on technological advancements, and it is anticipated that successful AI implementations, including those involving LLMs for complex text analysis and decision support in regulatory environments, become more widely reported as administrations gain experience and demonstrate tangible benefits.

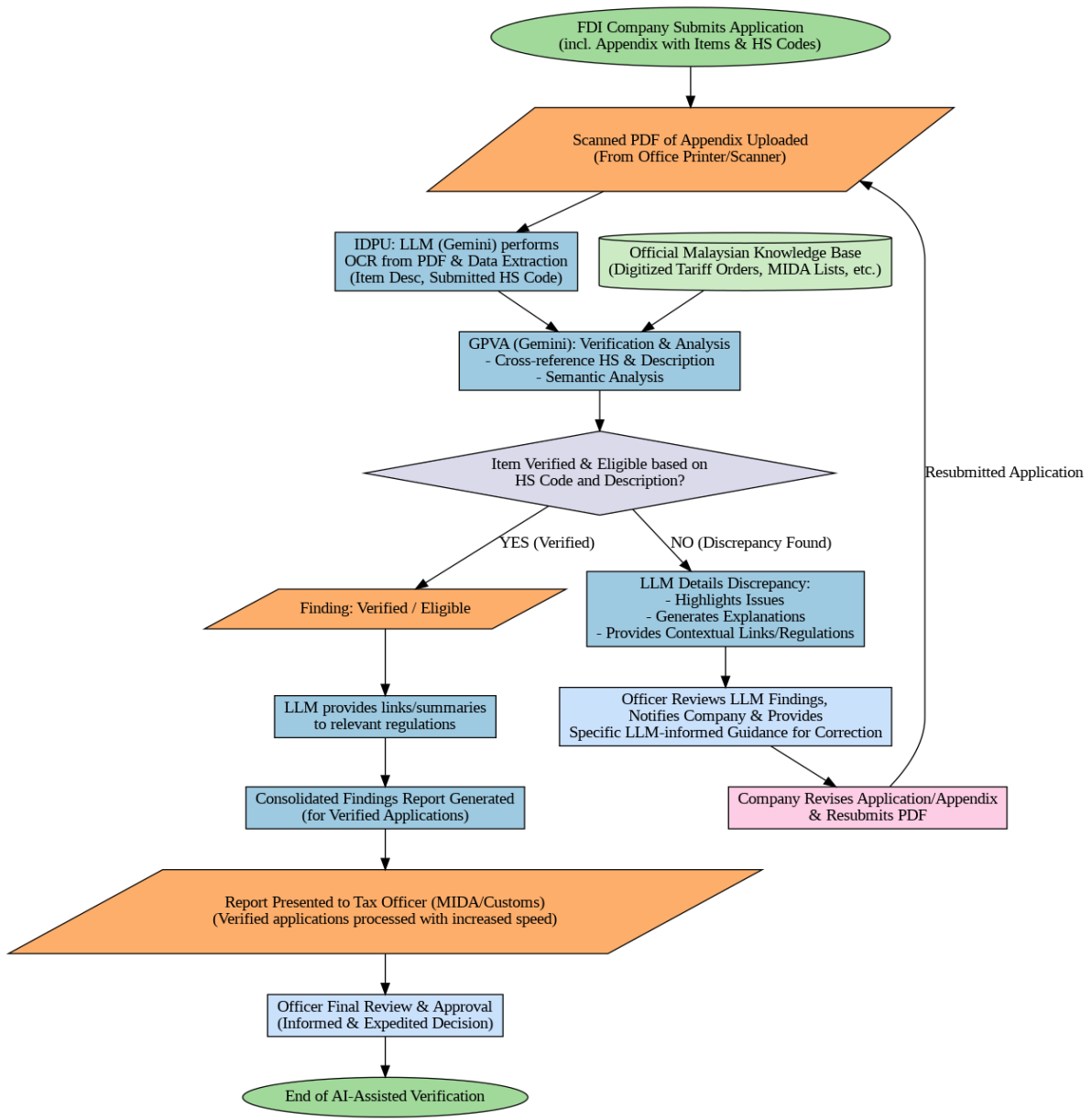

Figure 8: FDI tariff exemption verification flows with Intelligent Automation aid.

    Figure 8 flowchart visually outlines an automated system for processing Foreign Direct Investment (FDI) tariff exemption applications in Malaysia, leveraging intelligent automation with a feedback loop for corrections. The process begins when an FDI company submits the application, including an appendix with item descriptions and HS codes. This appendix, as a scanned PDF, is uploaded, and an Intelligent Document Processing Unit (IDPU) utilizing a Large Language Model (LLM) like Gemini performs Optical Character Recognition (OCR) from the PDF and extracts relevant data such as item descriptions and submitted HS codes. This extracted information is then verified by a Gemini-Powered Verification Assistant (GPVA) which cross-references it against an official Malaysian knowledge base of tariff orders and MIDA lists, performing semantic analysis for accuracy.

    Following this verification, a decision is made: if the item is verified and eligible, a "Verified / Eligible" finding is issued, the LLM provides links to relevant regulations, and a consolidated report is generated, leading to an expedited review and approval by the tax officer. However, if a discrepancy is found, the LLM details the issue, generates explanations, and provides contextual links/regulations. The tax officer then reviews these AI-generated findings, notifies the company, and offers specific, LLM-informed guidance for correction. The company revises and resubmits its application, which re-enters the verification flow, creating an iterative loop until compliance is achieved or a final decision is made. This system is designed to enhance precision, reduce the risk of non-compliance for FDI companies, expedite processing for compliant applications, and facilitate a clear correction pathway, ultimately aiming to improve Malaysia's ease of doing business in the manufacturing FDI sector.

### 2.7. Research Gap

The preceding review of literature reveals several key insights: Malaysia actively utilizes FDI incentives like tariff exemptions to drive manufacturing growth, a strategy underscored in national plans (Ministry of Investment, Trade and Industry, 2023) and supported by economic analyses (Abdullah & Zolkornain, 2021); effective tax administration is vital for this process, a principle established in public finance literature (Bird, 2015; Keen & Slemrod, 2017); yet, manual processes within public administration frequently present significant bottlenecks and risks (Bostrom & Heinen, 1977; Heeks, 2006). While OCR offers mature solutions for intelligent data capture from documents (Souza, Batista, & Silva, 2021), and advanced LLMs like Google's Gemini present powerful capabilities for interpreting complex regulatory texts and supporting decision-making within the RegTech space (Google AI, various publications; Zheng, Zhang, & Chen, 2021), a specific research gap exists. Current literature extensively covers these areas individually or in broader contexts, such as general AI adoption in public services (Al-Hashimi & Kim, 2021; Kuziemsky & Varpio, 2022) or global trends in customs modernization (Corteva, 2023). However, there is a discernible lack of focused investigation into how these distinct technologies OCR for initial document processing and a state-of-the-art, multimodal LLM like Google's Gemini (as understood in early 2025) can be synergistically integrated into a conceptual framework specifically designed to optimize the nuanced and critical process of tariff exemption verification for manufacturing FDI within the national tax administration. While national strategies call for public sector modernization through AI, detailed conceptual models addressing this particular high-impact administrative consideration by leveraging the unique strengths of Gemini are yet to be thoroughly explored and proposed. This research aims to address this gap by proposing such a framework, thereby contributing a targeted solution to a significant administrative and economic facilitation objective.

### 3. Methodology

#### 3.1. Research Paradigm and Approach

This study adopts a pragmatic research paradigm, focusing on practical application and outcomes by addressing a real-world problem within Malaysian tax administration. Pragmatism is suitable as the research aims to develop a useful conceptual framework intended to guide the implementation of an AI-driven solution (Aarons, Hurlburt, & Horwitz, 2011). The overall research approach is qualitative, allowing for an in-depth exploration and understanding of the complexities surrounding tariff exemption processes, the potential of AI technologies, and the administrative context within which the proposed system would operate. This qualitative approach aligns with methodologies described by Creswell and Poth (2016) and Miles, Huberman, and Saldaña (2018), emphasizing rich description and interpretation derived from a variety of textual data sources.

#### 3.2. Research Design

The research design is centered on the development of a conceptual framework. This design involves a systematic synthesis of existing knowledge, best practices, and technological capabilities to propose a novel system architecture tailored to the specific needs of Malaysian tax administration in managing FDI tariff exemptions. The process draws upon principles of framework development often seen in implementation science and information systems research, where understanding the context, components, and their interactions is crucial (Damschroder et al., 2009; Heeks, 2006). While not a full design science research project, it incorporates the design artifact creation aspect by proposing a detailed conceptual model. The design also implicitly uses elements of a case study approach (Yin, 2017) by focusing intensively on the Malaysian context, specifically the operations of RMCD and MIDA, as reflected in the analyzed documents and strategic plans.

#### 3.3. Data Sources and Collection

The primary data for this study is drawn from an extensive review of secondary sources, as comprehensively listed in the provided references. Data collection involved a meticulous selection and analysis of peer-reviewed academic journals, conference proceedings, books, official government publications and reports (e.g., from MITI, MIDA, MAMPU, MOSTI, RMCD), whitepapers from technology providers (e.g., Google AI, Google Research, Amazon Web Services / Microsoft Azure / Google Cloud), influential reports from international organizations (e.g., OECD, World Bank, WCO, UNCITRAL, UNDP), and frameworks from research institutes (e.g., The Alan Turing Institute, AI Now Institute, NIST). These documents provide foundational knowledge on FDI in Malaysia (Abdullah & Zolkornain, 2021; Athukorala, 2019), tax administration practices (Bird, 2015; Keen & Slemrod, 2017), challenges in public administration, OCR technology (Souza, Batista, & Silva, 2021), the architecture and capabilities of Large Language Models including Google's Gemini (Google AI, various publications; Bommasani et al., 2021; Zhao et al., 2023), AI in RegTech (Zheng, Zhang, & Chen, 2021), and AI adoption strategies both globally and within Malaysia. The collection focused on materials published up to early 2025 to ensure contemporary relevance, particularly concerning technological capabilities and national strategic directions.

### 3.4. Data Analysis and Framework Development

The analytical process for developing the conceptual framework is iterative and synthetic, integrating insights from the diverse data sources. The initial step involved a thematic analysis of the literature to delineate current administrative workflows for tariff exemptions in Malaysia, identify prevalent operational bottlenecks, and understand the risks associated with manual or sub-optimal processes. Concurrently, an analysis of technical literature, particularly concerning OCR (Smith, 2007; Souza, Batista, & Silva, 2021) and Google's Gemini LLM (Google AI, Google Research, various publications), was undertaken. This analysis specifically focused on Gemini's known strengths as of early 2025, such as its advanced reasoning, multimodal understanding (processing information from diverse document types that OCR would digitize), contextual awareness, extensive language support relevant to Malaysian documents, and its potential integration pathways via platforms like Google Suites or Google Cloud.

The core of the framework development involved mapping these identified technological capabilities onto the diagnosed administrative challenges. For instance, Gemini's advanced language understanding and reasoning capabilities were considered for their potential to interpret complex regulatory texts and official Malaysian directives related to HS Code classification (Chalkidis et al., 2020; Klanrit, Lee, & Kim, 2023), thereby assisting officers in making more accurate and consistent decisions. Its capacity to process and summarize large volumes of information was evaluated for reducing officer fatigue and expediting review times. The analytical process also involved designing the interrelationships between the OCR component (for data ingestion and digitization), the Gemini LLM (for analysis, verification, and recommendation), existing administrative databases, and the human customs/MIDA officers, conceptualizing an "AI as a co-pilot" interaction model (Amershi et al., 2019; Parasuraman, Sheridan, & Wickens, 2000). The framework's design explicitly aimed to enhance administrative efficiency, accuracy in HS code verification, control over exemption misuse, and overall transparency, contributing to an improved investment climate. This synthesis was guided by established principles of information systems design and socio-technical systems thinking (Bostrom & Heinen, 1977; Mumford, 2006) to ensure the proposed framework is both technologically sound and organizationally viable.

### 3.5. Ensuring Rigor and Trustworthiness

To ensure the rigor and trustworthiness of the proposed conceptual framework, several measures were employed. The primary strategy was triangulation, achieved by synthesizing information from a wide array of diverse sources, including academic literature, governmental reports, technical documentation, and international best practices (Creswell & Poth, 2016). This multifaceted review helped to validate findings and ensure a comprehensive understanding of the problem domain and potential solutions. The conceptual framework itself was developed systematically, with clear articulation of its components, their functions, and their interrelations, based on established theories and evidence from the literature. The explicit grounding of the framework's design in the documented capabilities of OCR and Google's Gemini (as of early 2025) further enhances its credibility as a plausible and relevant solution. The limitations of the proposed framework and the underlying methodology are also clearly acknowledged to provide a balanced perspective.

### 3.6. Ethical Considerations

In developing this framework, even though we're relying on secondary data and not directly interacting with people, we've carefully considered the ethical implications for the proposed AI system. This includes data privacy and security, especially for sensitive commercial and government data the system would handle. Our approach aligns with established data governance principles and, importantly, with Malaysia's own National Guidelines on AI Governance & Ethics (AIGE), launched in September 2024. The framework also inherently addresses the need for transparency and explainability in AI-assisted decisions. This is crucial for ensuring accountability and maintaining human oversight by officers. We've also acknowledged the potential impact on the workforce, such as changes in job roles and the need for upskilling, as a vital consideration for future implementation. Ultimately, this study champions the responsible development and deployment of AI within the public sector, guided by Malaysia's commitment to ethical AI.

### 3.7. Limitations of the Methodology

This study, while comprehensive in its review and synthesis, has inherent methodological considerations. Firstly, the reliance on publicly available secondary data means that insights into the nuanced, internal operational details of the national tax administration might be less granular than if primary data collection methods like interviews or direct observation were employed. Secondly, the proposed conceptual framework is, by nature, a theoretical construct. Its practical efficacy and implementation success can only be fully ascertained through subsequent empirical validation, pilot testing, and iterative refinement in a real-world setting. Thirdly, the depiction of Gemini's capabilities is based on information available up to early 2025; the rapidly evolving nature of LLMs means that specific functionalities and performance characteristics may evolve over time, potentially influencing some aspects of the proposed framework. Finally, the study focuses primarily on the technical and administrative aspects, and while ethical considerations are noted, a more extensive socio-economic impact assessment would be beneficial prior to full-scale deployment.

## 4. The Current Landscape of Tariff Exemption Administration in the National Context

### 4.1. Overview of Tariff Exemptions for Manufacturing FDI in Malaysia

Tariff exemptions represent a significant fiscal incentive utilized by Malaysia to attract and retain Foreign Direct Investment (FDI), particularly within the strategic manufacturing sector. As outlined in national economic agendas like the New Industrial Master Plan 2030 (Ministry of Investment, Trade and Industry, 2023), these exemptions are designed to reduce the initial investment costs and ongoing operational expenses for companies, thereby enhancing Malaysia's competitiveness and fostering industrial development (Abdullah & Zolkornain, 2021; Athukorala, 2019). The effective and efficient administration of these exemptions is therefore paramount, not only for delivering the intended economic benefits but also for maintaining the integrity of the nation's fiscal regime and ensuring a transparent and predictable environment for investors. The Royal Malaysian Customs Department (RMCD) and the Malaysian Investment Development Authority (MIDA) are central to this administrative ecosystem, overseeing the application, verification, and approval processes for these critical incentives.

### 4.2. Key Administrative Bodies and Their Roles: RMCD and MIDA

The administration of tariff exemptions for manufacturing FDI in Malaysia is primarily managed by two key governmental bodies: The Royal Malaysian Customs Department (RMCD) and the Malaysian Investment Development Authority (MIDA). MIDA, as the principal national investment promotion agency, often plays the initial role in facilitating FDI, providing information on available incentives, and guiding investors through the application process (Malaysian Investment Development Authority, various years). They are typically involved in assessing the eligibility of projects and companies for various fiscal incentives, including tariff exemptions, based on criteria aligned with national development goals. Subsequently, the RMCD is responsible for the customs-specific aspects of these exemptions. This includes the critical tasks of verifying the classification of imported raw materials, components, and machinery against the Harmonized System (HS) codes, ensuring compliance with the terms of the exemption, processing customs declarations, and conducting post-importation audits (Royal Malaysian Customs Department, various years). The collaboration and information exchange between MIDA and RMCD are crucial for a seamless exemption process, though inter-agency coordination can itself present challenges in public administration.

### 4.3. Current Workflow for Tariff Exemption Application and Verification

While specific internal workflows can vary and evolve, the general process for obtaining and utilizing tariff exemptions for manufacturing FDI involves several key stages, often characterized by substantial documentation and existing checks. Typically, an investing company submits an application detailing its project, the goods for which exemption is sought, their intended use, and estimated import values. This application is supported by a considerable volume of documentation, such as business registration details, manufacturing licenses, lists of raw materials and machinery, technical specifications, invoices, and bills of lading. Officers then undertake a verification process. This involves scrutinizing the submitted documents for completeness and authenticity, assessing the eligibility of the investor and the specific goods against prevailing policies and customs regulations, and, crucially, verifying the correct HS Code classification for each item. This classification determines the applicable standard tariff rate and thus the value of the exemption. This stage often involves cross-referencing of product descriptions with extensive tariff schedules, legal notes, and official customs directives. Approval, if granted, comes with conditions and reporting requirements that necessitate ongoing compliance monitoring.

### 4.4. Identified Challenges and Bottlenecks in the Current System

The current system for administering tariff exemptions, while foundational, presents several significant areas for optimization and operational considerations that can influence efficiency, precision, and the overall experience for investors. These considerations inherently set the stage for exploring how advanced AI capabilities, such as those offered by Google's Gemini, could provide targeted solutions. A primary area for enhancement is the rigorous application of the General Interpretative Rules (GIRs) in a strict hierarchical order. For instance, GIR 1 dictates that classification is determined by heading terms and relative Section/Chapter Notes, which frequently contain exclusionary clauses. Optimizing the identification and application of these Notes, or interpreting the scope of a heading, can lead to more precise classifications even before subsequent GIRs are considered. The General Interpretative Rules (GIR) are a set of six principles used to ensure the uniform classification of goods under the Harmonized System (HS) for tariff purposes. GIR 1 establishes the primary basis for classification: the terms of the headings and any relative Section or Chapter Notes. If classification cannot be determined by GIR 1, subsequent rules are applied sequentially. GIR 2(a) extends classification to incomplete/unfinished goods if they possess the essential character of the complete/finished article, and to unassembled/disassembled goods. GIR 2(b) addresses mixtures and combinations of substances, stating that they should be classified according to the principles of GIR 3 if they consist of more than one material. GIR 3 provides a hierarchy for goods classifiable under multiple headings: GIR 3(a) directs to the most specific description; if that's not determinative, GIR 3(b) classifies based on the material or component that gives the goods their essential character; and if neither (a) nor (b) applies, GIR 3(c) selects the heading that occurs last in numerical order. Should these rules still not suffice, GIR 4 allows for classification under the heading of goods to which they are most akin. GIR 5 then specifically addresses the classification of cases, boxes, and similar containers (GIR 5(a)) and packing materials and containers (GIR 5(b)), generally classifying them with the goods they contain under certain conditions. Finally, GIR 6 mandates that the classification of goods in the subheadings of a heading shall be determined by the terms of those subheadings and any related Subheading Notes, applying GIRs 1 to 5, mutatis mutandis, on the understanding that only subheadings at the same level are comparable. A primary consideration lies in the nuances of regulatory interpretation and HS Code classification. National Customs Orders and Regulations are extensive and subject to periodic updates, requiring officers to possess comprehensive and current knowledge. The linguistic intricacies of tariff nomenclature and the nuanced distinctions between HS codes make classification a cognitively intensive task, with potential for variability, an aspect broadly recognized in international trade. This dynamic is a key area where Gemini's advanced language understanding and reasoning could assist in interpreting regulatory texts and suggesting precise classifications. Secondly, the volume of documentation and data involved in applications is substantial. Officers actively review and extract relevant information from numerous, often lengthy and varied, document formats. Optimizing this data processing can enhance efficiency, redirect skilled officers to higher-value analytical tasks, and mitigate the potential for data entry variations or oversights. OCR technology can address the digitization aspect, and Gemini's multimodal capabilities and capacity to understand and summarize content from diverse documents could further streamline this data extraction and preliminary analysis. Thirdly, existing processes inherently present opportunities for accelerating turnaround times and optimizing administrative flow. The time taken to thoroughly verify each application can be a factor, influencing project timelines for FDI companies and potentially affecting investment attractiveness. Reducing administrative load through intelligent assistance and expediting verification steps with AI are potential benefits that Gemini could offer. Fourthly, the system presents opportunities for enhancing uniformity and managing potential non-alignment of exemptions. Variations in interpretation among officers or across different processing centers can influence rule application. Moreover, the volume and complexity of transactions offer an opportunity to enhance the detection of and deterrence against misclassification or other forms of non-compliance with exemption conditions through intelligent auditing. An AI system like Gemini could enhance uniformity by providing standardized interpretations and strengthen risk assessment by identifying anomalies or patterns indicative of potential non-alignment. Finally, data management and knowledge sharing within and between relevant administrative bodies present opportunities for optimization. Information within disparate systems or paper files can influence efficient cross-referencing and comprehensive oversight. An AI system could contribute by creating a more structured and accessible knowledge base derived from processed applications and regulatory documents, facilitating better-informed decision-making. These multifaceted considerations underscore the opportunity for strategic technological intervention to modernize tariff exemption administration.

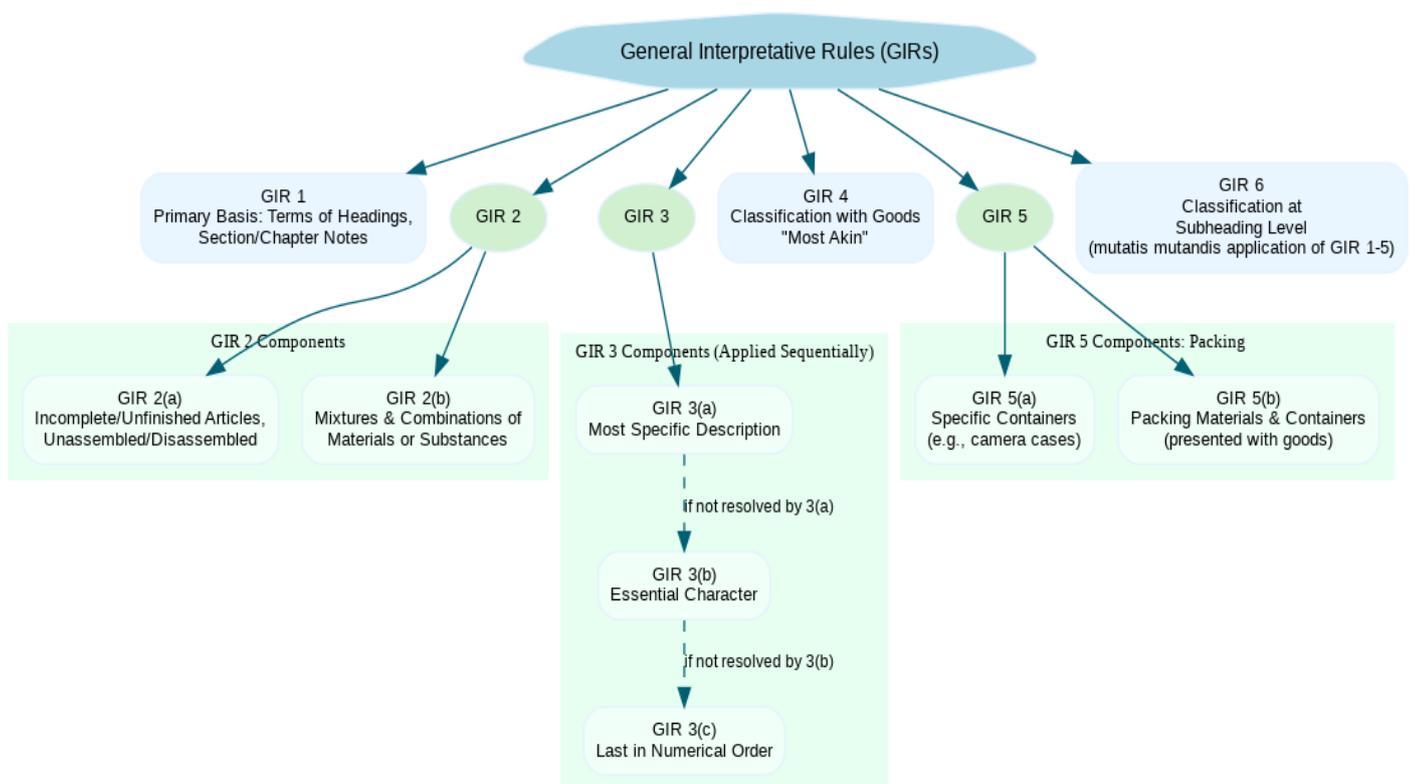

Figure 9: General Interpretative Rules for HS Code Classification

Referring to figure 9, the General Interpretative Rules (GIR) are a set of six principles used to ensure the uniform classification of goods under the Harmonized System (HS) for tariff purposes. GIR 1 establishes the primary basis for classification: the terms of the headings and any relative Section or Chapter Notes. If classification cannot be determined by GIR 1, subsequent rules are applied sequentially. GIR 2(a) extends classification to incomplete/unfinished goods if they possess the essential character of the complete/finished article, and to unassembled/disassembled goods. GIR 2(b) addresses mixtures and combinations of substances, stating that they should be classified according to the principles of GIR 3 if they consist of more than one material. GIR 3 provides a hierarchy for goods classifiable under multiple headings: GIR 3(a) directs to the most specific description; if that's not determinative, GIR 3(b) classifies based on the material or component that gives the goods their essential character; and if neither (a) nor (b) applies, GIR 3(c) selects the heading that occurs last in numerical order. Should these rules still not suffice, GIR 4 allows for classification under the heading of goods to which they are most akin. GIR 5 then specifically addresses the classification of cases, boxes, and similar containers (GIR 5(a)) and packing materials and containers (GIR 5(b)), generally classifying them with the goods they contain under certain conditions. Finally, GIR 6 mandates that the classification of goods in the subheadings of a heading shall be determined by the terms of those subheadings and any related Subheading Notes, applying GIRs 1 to 5, *mutatis mutandis*, on the understanding that only subheadings at the same level are comparable.

## 5. Conceptual OCR/Gemini LLM-Powered System for National Tax Administration

This chapter outlines a conceptual framework for an Optical Character Recognition (OCR) and Google Gemini Large Language Model (LLM)-powered system. It's designed to support Malaysian tax administration, specifically the Royal Malaysian Customs Department (RMCD) and the Malaysian Investment Development Authority (MIDA), in managing tariff exemptions for Foreign Direct Investment (FDI).

### 5.1. Design Philosophy: AI as a "Co-pilot" for National Tax Administration Officers

The core design philosophy for the proposed system is AI as a "co-pilot," where technology augments and enhances the capabilities of human officers rather than replacing them. This approach, drawing from guidelines for human-AI interaction (Amershi et al., 2019; Shneiderman, 2020), ensures that officers retain final decision-making authority while being empowered by the advanced analytical and information processing strengths of Google's Gemini LLM. The system aims to function as a sophisticated decision-support tool (Benbasat & Lim, 2000), helping officers navigate complexity, reduce cognitive load, and improve the consistency and accuracy of their work, thereby maintaining their situation awareness (Endsley, 2017). By leveraging Gemini's capabilities (as understood in early 2025, detailed in Google AI and Google Research publications) within a human-centered design, the system intends to create a collaborative synergy between human expertise and artificial intelligence, aligning with principles of joint cognitive systems (Hollnagel & Woods, 2005). This philosophy emphasizes trust, transparency, and the ultimate goal of enabling officers to perform their duties more effectively and efficiently.

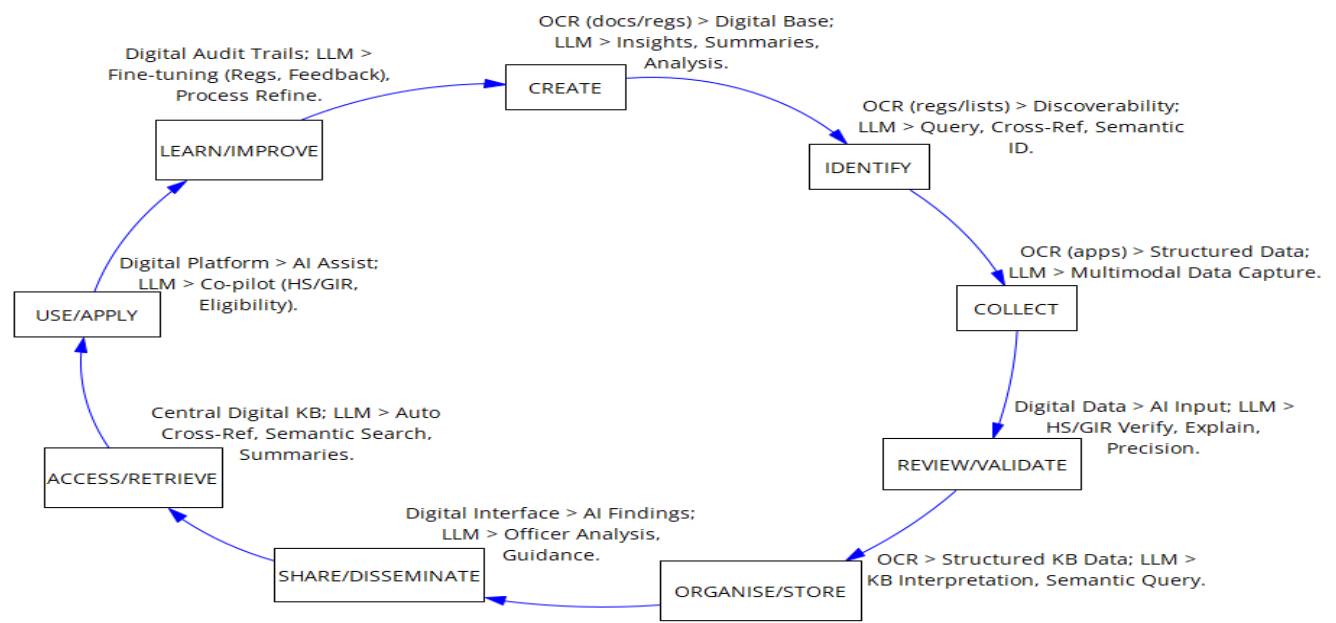

Figure 10: Improvised APQC's (American Productivity & Quality Center) classic Knowledge Flow Process

Figure 10 illustrates a modernized APQC 9-step knowledge management cycle, depicted as a continuous circular flow, where each stage from CREATE through to LEARN/IMPROVE is significantly enhanced by embedded Digitalization and Artificial Intelligence (D&AI) technologies, particularly Optical Character Recognition (OCR) and Large Language Models (LLMs). As visualized for a context likely relevant to Malaysian trade or regulatory processes (e.g., involving "HS/GIR" tariff classifications), these D&AI tools are shown to transform traditional knowledge activities by enabling automated data capture, intelligent analysis, semantic search, AI-assisted decision-making, and continuous process refinement, ultimately aiming for a more efficient, accurate, and intelligent knowledge flow.

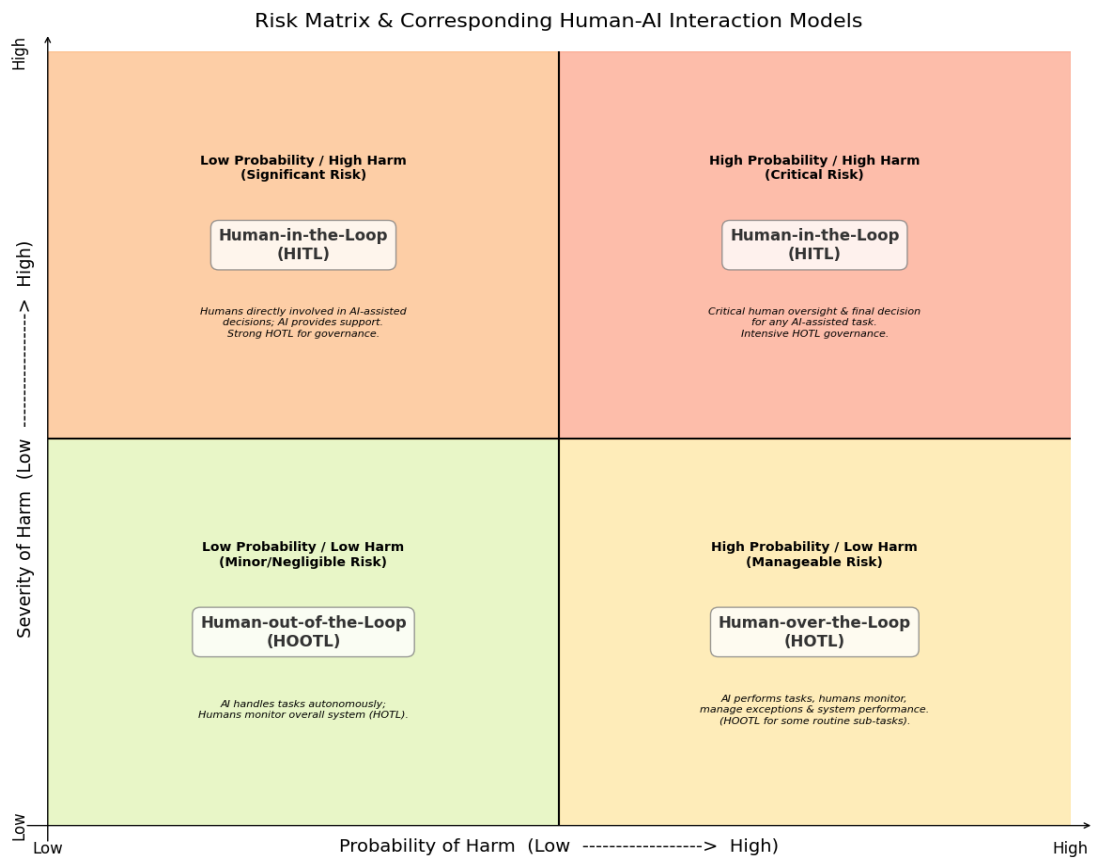

Figure 11: Risk quadrant Matrix and corresponding human-AI interaction Models

The 4-quadrant risk matrix on figure 11, which plots the "Severity of Outcome" against the "Likelihood of Outcome," is a tool used to categorize and prioritize potential scenarios. This matrix also helps in conceptualizing the appropriate level of human oversight when AI systems are employed, suggesting models like Human-out-of-the-Loop (HOOTL) for scenarios with lower impact/likelihood, Human-over-the-Loop (HOTL) for manageable ones, and crucial Human-in-the-Loop (HITL) for significant or critical outcomes. In the context of "Intelligent Automation for Investment Facilitation," this matrix is highly relevant because the tariff exemption process inherently involves potential scenarios such as "non-optimized exemption utilization" and "non-compliance," which can lead to significant financial, legal, or reputational considerations for both companies and the national tax administration. The proposed OCR-LLM system aims to "enhance the precision of critical verification tasks" and "drastically reduce the potential for non-compliance," effectively lowering the likelihood of these outcomes. Given the potential for high severity in tariff exemption outcomes, these considerations could fall within the upper, more critical quadrants of the matrix. Therefore, the system's design, which empowers tax officers with AI as a "co-pilot" (a Human-in-the-Loop approach), is particularly suitable for managing these high-stakes scenarios, aiming to shift potential outcomes from a higher likelihood/severity quadrant to a state where the likelihood is reduced while human oversight ensures the higher severity potential is carefully managed.

**5.2. System Architecture and Core Modules**

The proposed system comprises two primary interconnected modules designed to streamline the tariff exemption verification process from document ingestion to officer decision support.

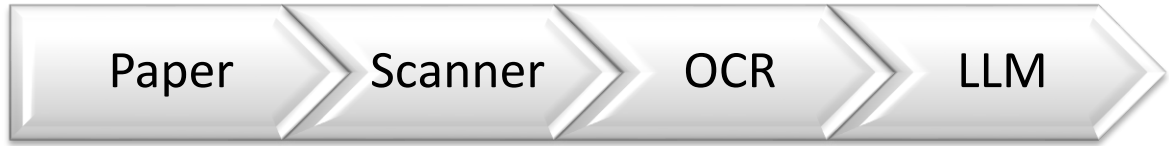

Figure 12: Flow-chart process of transformation of digitalization of LLM input data

**Module 1: Intelligent Document Processing Unit (IDPU)**

The Intelligent Document Processing Unit (IDPU) serves as the initial gateway for all application-related documents. This module is fundamentally OCR-based, employing advanced OCR technologies (Souza, Batista, & Silva, 2021; Smith, 2007) to convert various forms of submitted documentation such as application forms, invoices, bills of lading, technical specifications, and certificates into machine-readable text and structured data. Beyond basic digitization, the IDPU incorporate intelligent features to classify document types, identify key information fields, and extract relevant data points (e.g., applicant details, item descriptions, claimed HS codes, quantities, values). This accurately extracted and structured data is crucial, as it forms the primary input for the subsequent analysis by the Gemini-powered module, ensuring that Gemini operates on clean and well-organized information.

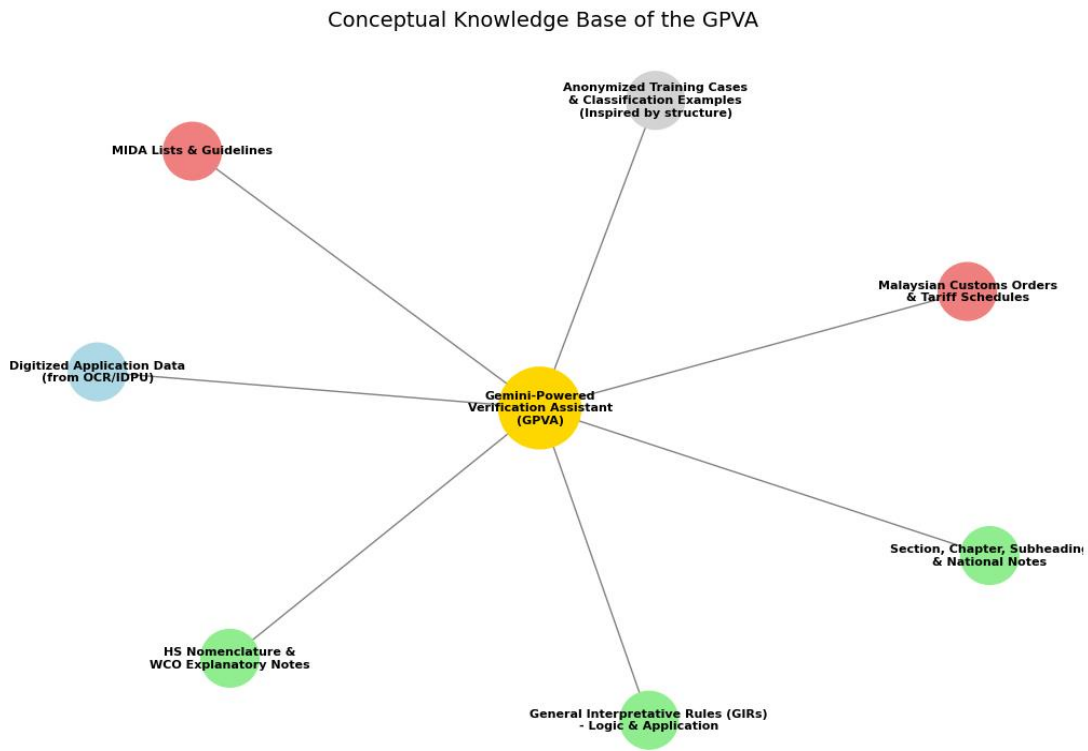

Figure 13: Conceptual knowledge base of the GPVA

**Module 2: Gemini-Powered Verification Assistant (GPVA) – For Officer Use**

The Gemini-Powered Verification Assistant (GPVA) is the core analytical engine of the system, designed for direct use by RMCD and MIDA officers. It leverages Google's Gemini LLM's advanced capabilities to provide comprehensive verification support. Its key functionalities include:

Automated cross-referencing and knowledge base querying: The GPVA take the structured data from the IDPU (e.g., item descriptions, quantities, claimed HS codes) and automatically cross-reference it against an integrated knowledge base. This knowledge base would comprise official Malaysian customs orders, tariff schedules, lists of eligible/ineligible items, MIDA's guidelines, and other relevant regulatory documents, which can be regularly updated. Gemini's analytical capabilities (Google AI, various publications; Zhao et al., 2023) be employed to perform complex queries and pattern matching within this extensive dataset, far exceeding the speed and scope of manual checks.

Semantic analysis for classification and eligibility: A critical function of the GPVA be its Gemini-driven semantic analysis. This involves using Gemini's sophisticated natural language understanding (NLU) and reasoning abilities (Brown et al., 2020; Hendrycks et al., 2021) to interpret the nuanced descriptions of imported goods. It compares these descriptions against the detailed specifications in the Harmonized System (HS) nomenclature and local Malaysian regulatory texts (Chalkidis et al., 2020; Klanrit, Lee, & Kim, 2023) to verify the appropriateness of the claimed HS code or suggest more accurate alternatives. Furthermore, it analyzes item eligibility based on the specific criteria for tariff exemptions, flagging items that may not qualify. The model could be fine-tuned on Malaysian regulatory and customs language (Gururangan et al., 2020) to enhance its domain-specific accuracy.

Discrepancy highlighting and explainable outputs: The GPVA be designed to proactively highlight potential discrepancies, ambiguities, non-listed HS codes, or items deemed ineligible. Crucially, it provides officers with clear, concise explanations for these flags, generated or supported by Gemini's reasoning capabilities (Adadi & Berrada, 2018; Guidotti et al., 2018; Miller, 2019). This aligns with the principles of explainable AI (XAI), enabling officers to understand the basis of the AI's suggestions and build trust in the system (Wang et al., 2021). The goal is not black-box decisioning but transparent assistance (Rai, 2020).

Contextualized information and regulatory linkage: For items where verification is complex or requires deeper scrutiny, the GPVA provide officers with direct hyperlinks to the relevant sections of underlying laws, regulations, customs rulings, or official lists. Gemini can further enhance this by offering summaries or contextual explanations of these legal texts (Zheng, Zhang, & Chen, 2021), making it easier for officers to quickly grasp the pertinent rules and confirm their assessments.

Semantic analysis for classification and eligibility: This involves using Gemini's NLU to interpret item descriptions against HS nomenclature, and critically, to apply the logic of the General Interpretative Rules (GIRs). For instance, when analyzing an 'unassembled bicycle', Gemini could identify the applicability of GIR 2(a) and prompt for or verify its 'essential character'. Similarly, for goods potentially classifiable under multiple headings, it could flag the need to apply GIR 3, perhaps suggesting the most specific description (GIR 3(a)) or the component providing essential character (GIR 3(b)). A key strength would be its ability to parse and apply exclusionary or inclusionary Section and Chapter Notes, such as those determining the classification of a plastic toy away from Chapter 39 into Chapter 95.

**5.3. Operational Workflow: How Officers Interact with the Gemini-Powered GPVA**

The operational workflow is designed to integrate the GPVA seamlessly into the daily tasks of administrative officers, enhancing their existing processes. Application Submission & Digitization: Applicants submit their tariff exemption applications and supporting documents (potentially through an online portal or as scanned documents). The IDPU automatically processes these documents, extracts relevant data, and indicates any missing information. Case Assignment & Initial Review by GPVA: Digitized applications are assigned to an officer. The GPVA has already performed a preliminary analysis on the case, including the automated cross-referencing and semantic analysis. Officer Interaction with GPVA Interface: The officer accesses the case via a dedicated interface. This interface presents a consolidated view of the application, with key data highlighted. The GPVA's findings such as suggested HS code verifications, eligibility checks, highlighted considerations, and confidence score be clearly displayed alongside the original document snippets. Guided Verification & Decision Making: Officers use the GPVA's outputs as a guide. They can review indicated items, explore the AI-generated explanations, access linked regulatory texts (potentially summarized by Gemini), and examine the evidence. Officers can query the GPVA for further clarification or to test alternative scenarios. The system supports naturalistic decision-making processes by providing information in an intuitive and actionable manner. Officer Adjudication & Audit Trail: The officer makes the final decision on each item and the overall application, with the ability to override AI suggestions by providing justification. All interactions, AI suggestions, officer decisions, and justifications are logged, creating a comprehensive audit trail crucial for transparency and accountability. Feedback Loop & Continuous Improvement: The system could incorporate a feedback mechanism where officers can rate the usefulness of AI suggestions, which, over time (and with appropriate safeguards and aggregation), can be used to further refine the underlying models and the knowledge base. This workflow ensures that the officer remains central to the process, using the GPVA as a powerful assistant to navigate complexity and enhance decision quality.

**5.4. Data Management, Security, and Ethical Considerations for using Gemini within Government**

The deployment of an OCR/Gemini-powered system within Malaysian tax administration necessitates robust data management, security protocols, and adherence to ethical AI principles. Data Management and Security: All data processed by the system, including sensitive commercial information from FDI applicants and official regulatory data, must be managed in accordance with Malaysian data governance laws and best practices (Puthucheary & Sivanathan, 2022; AlRyalat, Al-Hawari, & Al-Omari, 2021). If leveraging Google Cloud infrastructure for hosting Gemini and associated services, data residency options ensuring data remains within Malaysia (or approved jurisdictions) must be utilized, as per cloud provider documentation (Amazon Web Services / Microsoft Azure / Google Cloud, various whitepapers). Strict access controls, role-based permissions, encryption of data at rest and in transit, and regular security audits are paramount, aligning with frameworks like the NIST AI Risk Management Framework (NIST, various years) and general cybersecurity principles for AI (Rowland

& Podnar, 2021; Comiter, 2019). Comprehensive logging and monitoring essential to detect and respond to any security incidents or unauthorized access attempts.

Ethical Considerations and Responsible AI: The use of Gemini must align with established ethical guidelines for trustworthy AI (European Commission's High-Level Expert Group on AI, 2019; Floridi et al., 2018; Dignum, 2019). This includes ensuring fairness (Aggarwal, 2018, on biases), transparency in how the AI arrives at suggestions (as discussed with XAI), human oversight, and accountability for final decisions. The system should be designed to avoid algorithmic bias that could disproportionately affect certain types of applications or investors. Regular assessments for potential biases and performance drift will be necessary. Specific attention must be paid to Malaysia's developing AI regulatory landscape and legal perspectives on AI ethics (Ariffin & Ahmad, 2023; Jobin, Ienca, & Vayena, 2019). The system should not engage in manipulative "nudging" (Susser, Roessler, & Nissenbaum, 2019) but rather provide clear, objective support. Considerations of digital sovereignty (Pohle & Thiel, 2020; Wagner, 2018) and the governance of algorithmic systems by global platforms within national contexts (Katzenbach & Ulbricht, 2019) will also inform policy and oversight. The overall design must uphold principles of value-sensitive design (van den Hoven, Vermaas, & van de Poel, 2015), ensuring the technology serves public value and maintains public trust.

## 6. Benefits, Strategic Considerations, and Broader Implications

The proposed conceptual framework for an Optical Character Recognition (OCR) and Google Gemini Large Language Model (LLM)-powered system offers significant potential for transforming tariff exemption administration in Malaysia. This chapter explores the direct benefits for the Malaysian tax administration and Foreign Direct Investment (FDI) companies, its impact on the national investment climate, and critically examines the implementation challenges and mitigation strategies associated with deploying such an advanced AI solution.

### 6.1. Direct Benefits for Malaysian Tax Administration (RMCD & MIDA) through Gemini Implementation

The integration of Google's Gemini LLM into the tariff exemption workflow promises substantial direct benefits for the national tax administration. A primary advantage lies in enhanced operational efficiency. Gemini's capability for rapid, large-scale analysis of complex regulatory texts and application documents, as highlighted by Google AI and Google Research publications, can significantly reduce the time officers spend on manual verification and cross-referencing tasks. This accelerated processing, from document ingestion via OCR to AI-assisted verification, can lead to faster turnaround times for applications, thereby optimizing resource allocation within these agencies (De Witte & Geys, 2011; Kuziemsky & Varpio, 2022).

Furthermore, the system is anticipated to bring about a notable improvement in accuracy and consistency. Gemini's advanced language understanding and reasoning capabilities (as known in early 2025) can aid in the precise interpretation of intricate customs nomenclature and eligibility criteria, minimizing human errors that may arise from fatigue or subjective judgment (Keen & Slemrod, 2017). This heightened accuracy directly addresses the documented complexities in customs classification, such as the consistent application of the six General Interpretative Rules, the correct interpretation of often nuanced Section and Chapter Notes (which can alter prima facie classifications), and ensuring uniformity in decisions that could otherwise vary due to subjective interpretation of 'essential character' or 'specific description. By providing standardized interpretations and flagging discrepancies with AI-generated explanations, the system can ensure more uniform application of rules across different officers and cases, contributing to fairer and more reliable outcomes. This increased accuracy also translates to enhanced control and risk management, as the system can potentially identify patterns or anomalies indicative of non-compliance or misuse more effectively than purely manual reviews. Ultimately, the adoption of such a system signifies a crucial step towards the modernization of public administration (Barcevičius, Cibaitė, & Codagnone, 2019; Mehr, 2017), aligning with broader digital transformation goals and improving the quality of government services (El-Haddadeh, Weerakkody, & Osmani, 2019).

### 6.2. Positive Outcomes for FDI Companies

The implementation of an efficient and accurate OCR/Gemini-powered system for tariff exemptions is expected to yield significant positive outcomes for FDI companies operating in or considering Malaysia. The most immediate benefit will be reduced processing times for tariff exemption applications. Faster approvals mean companies can import necessary raw materials, components, and machinery more quickly, accelerating project timelines and enabling quicker commencement of manufacturing operations. This efficiency directly translates into potential cost savings and improved cash flow management for businesses.

Moreover, the system can lead to greater predictability and transparency in the exemption process. Consistent application of rules and clearer communication (potentially facilitated by AI-generated explanations or status updates) can reduce uncertainty for investors (Felbermayr, Teti, & Yalcin, 2021). A more streamlined and understandable process reduces the administrative burden on companies, freeing up their resources to focus on core business activities rather than navigating complex bureaucratic procedures. This improved experience can enhance investor confidence and satisfaction, making

Malaysia a more attractive location for foreign investment, a key consideration often highlighted by international business organizations (International Chamber of Commerce, various publications).

### 6.3. Impact on Malaysia's Investment Climate and Ease of Doing Business

The benefits accruing to tax administration and FDI companies will collectively contribute to a more positive impact on Malaysia's overall investment climate and its ranking in global ease of doing business indices (World Bank, various years). An efficient, transparent, and technologically advanced customs and tax administration system is a critical determinant of a country's attractiveness to foreign investors. By significantly improving a key aspect of the FDI facilitation process, Malaysia can further enhance its reputation as a pro-business destination committed to leveraging cutting-edge technology for good governance. This modernization aligns with the strategic objectives of national plans like the New Industrial Master Plan 2030 (Ministry of Investment, Trade and Industry, 2023) which aim to draw high-quality FDI by creating a conducive ecosystem. The successful implementation of such an AI system can signal Malaysia's leadership in digital government within the region, fostering greater economic growth and development (Abdullah & Zolkornain, 2021; Athukorala, 2019). This contributes directly to the creation of public value by making public services more responsive and supportive of economic activity (Moore, 1995; Bryson, Crosby, & Bloomberg, 2014; Twizeyimana & Andersson, 2019).

### 6.4. Implementation Challenges and Mitigation Strategies for a Gemini-based solution

Despite the promising benefits, the implementation of a sophisticated AI solution like the proposed OCR/Gemini-powered system presents several challenges that require careful consideration and proactive mitigation strategies.

First, customizing and fine-tuning Gemini for specific Malaysian regulatory language and customs nomenclature is a critical hurdle. While powerful, general LLMs like Gemini may require significant adaptation to accurately interpret the highly specialized, often nuanced, terminology found in Malaysian legal and customs documents. Mitigation involves employing advanced fine-tuning techniques (Howard & Ruder, 2018; Gururangan et al., 2020), developing domain-specific prompts, and curating comprehensive datasets of Malaysian regulations for adaptation (Liao et al., 2023). Collaboration with legal and customs experts will be essential in this iterative process, drawing upon existing research in Legal NLP (Chalkidis et al., 2020; Klanrit, Lee, & Kim, 2023).

Second, managing dependencies on Google's platform, including updates and API changes, poses a strategic challenge. Relying on a third-party platform introduces risks related to vendor lock-in, unexpected modifications to service offerings, or changes in API functionalities that could impact the system's stability. To mitigate this, the Malaysian government should seek robust Service Level Agreements (SLAs). Crucially, to further manage these dependencies, consideration should be given to architectural best practices, such as developing an 'anti-corruption layer' or an abstraction layer within the system. This software engineering approach would serve to decouple the core Malaysian administrative system from direct, hard-coded dependencies on specific Google Gemini API structures. Such a layer would manage interactions with the LLM, making it easier to adapt to API updates from Google or, in the long term, even facilitate migration to alternative LLM solutions if strategically necessary, thereby enhancing system resilience and future-proofing the investment against unforeseen platform shifts. The government should also maintain open communication with Google regarding their product roadmap and advocate for long-term support commitments. Understanding the broader dynamics of digital technology governance (Flyverbom, Deibert, & Matzner, 2019) is also pertinent.

Third, ensuring robust data governance and security within the Google Cloud ecosystem, if utilized, is paramount. The system will handle sensitive government and commercial data, necessitating stringent adherence to Malaysian data protection laws (Puthucheary & Sivanathan, 2022) and international best practices. Mitigation strategies, as outlined in Chapter 5, include strong encryption, strict access controls, regular security audits, and leveraging Google Cloud's security features and data residency options to ensure data sovereignty. Comprehensive data governance frameworks specific to this AI application will be needed (AlRyalat, Al-Hawari, & Al-Omari, 2021; Janssen, van der Voort, & Wahyudi, 2017).

Fourth, the cost implications of scaled Gemini usage, even with potential government agreements, must be carefully managed. The computational resources required for sophisticated LLMs can lead to significant operational costs, particularly with high volumes of complex queries. Mitigation involves negotiating favorable pricing, optimizing the design of queries and workflows for efficiency, implementing robust usage monitoring and quota systems, and conducting a thorough, ongoing cost-benefit analysis to ensure sustained value.

Finally, the need for specialized training for officers on using Gemini-powered tools is crucial for successful adoption. Officers must be trained not only on the technical operation of the system but also on how to interpret AI outputs critically, understand its limitations, and effectively collaborate with the AI "co-pilot" (Amershi et al., 2019). Comprehensive change management programs (Fullan, 2006), user-friendly interface design, and continuous learning support will be essential to build trust, ensure user acceptance (Venkatesh et al., 2003), and effectively integrate this new technology into established work practices (Nonaka, 1994; Shneiderman, 2020). This aligns with broader needs for upskilling the public sector workforce for the digital age (OECD, various studies).

## 7. Conclusion

This chapter concludes the research by recapitulating its core focus on leveraging Google's Gemini Large Language Model (LLM) as a catalyst for modernizing tax administration in the national context. It summarizes the key findings regarding the transformative potential of integrating Optical Character Recognition (OCR) with Gemini for tariff verification, offers specific recommendations for key stakeholders, acknowledges the study's considerations, and proposes avenues for future research.

### 7.1. Recapitulation of Research: Google's LLM as a Catalyst for Modern Tax Administration in Malaysia

This research set out to address the nuances and opportunities for optimization inherent in the administration of tariff exemptions for Foreign Direct Investment (FDI) within national tax bodies. Recognizing the nation's strategic investment in advanced AI capabilities, including access to Google's Gemini LLM, this study investigated how these technologies could be harnessed. The central aim was to propose a conceptual framework for an OCR-powered system, integrated with Gemini, specifically designed to enhance the precision, efficiency, and transparency of verifying tariff exemption applications. By synthesizing insights from literature on AI in public administration, tax and customs modernization, and the technical capabilities of OCR and advanced LLMs like Gemini, this study has laid out a vision for a digitally augmented administrative process, aligning with national economic goals and digital transformation agenda. Mitigation involves curating comprehensive datasets of national regulations for adaptation. This dataset should ideally include not only formal tariff schedules and legal notes but also practical examples and common considerations for HS classification. Fine-tuning Gemini on such practical examples would equip it to better understand the application of rules, beyond just their literal text.

### 7.2. Key Findings: The Transformative Potential of OCR Integrated with Gemini for Tariff Verification

The key finding of this research is the significant transformative potential that an integrated OCR and Google Gemini LLM system holds for the national tariff verification process. The proposed conceptual framework demonstrates that such a system can move beyond simple automation to provide intelligent assistance to administrative officers. The OCR component addresses the initial opportunities for optimization in data entry from diverse documentation. More profoundly, Google's Gemini, with its advanced natural language understanding, reasoning, and (as of early 2025) multimodal capabilities, offers the potential to significantly improve the precision of HS Code classification by interpreting complex regulatory texts and item descriptions. It can expedite the verification process through rapid cross-referencing and analysis, enhance consistency in decision-making by providing standardized interpretations, and improve officer productivity by acting as an AI "co-pilot". The ability of Gemini to provide explainable outputs is crucial for maintaining officer oversight and building trust, thereby fostering a collaborative human-AI environment. This synergy promises not only to streamline administrative workflows but also to enhance the integrity and effectiveness of the tariff exemption scheme, contributing to a more efficient public service (De Witte & Geys, 2011).

### 7.3. Recommendations

Based on the findings and the proposed conceptual framework, the following recommendations are offered to key stakeholders to translate this potential into tangible benefits. For the national tax authority, several actions are advised. Firstly, it is recommended that the tax authority collaboratively develop a strategic plan focused on leveraging existing government access to Google's Gemini LLM for optimizing tariff exemption processing. This plan should align with national AI and public sector modernization strategies and must detail clear objectives, timelines, and performance metrics for AI integration. Following strategic planning, the initiation of focused pilot programs is crucial. These programs should utilize Gemini for specific tasks within the tariff verification workflow, particularly for HS Code verification and regulatory text interpretation, providing valuable insights into practical considerations, system performance, and user acceptance in a controlled environment. To this end, it is recommended that pilot programs incorporate specific Key Performance Indicators (KPIs) to quantitatively assess the impact of the Gemini-powered system. These should include, but not be limited to: (i) average percentage reduction in processing time per tariff exemption application, particularly for HS Code verification tasks, aiming for targets suggested by preliminary estimates (e.g., the potential 5.0 minutes per 300 codes for LLM-assisted checks versus 450.0 minutes manually); (ii) measurable improvement in the accuracy rate of initial HS code classifications compared to baseline manual processes; (iii) reduction in the number of applications requiring secondary review or correction due to classification errors; (iv) officer time saved on document review and information extraction, allowing reallocation to higher-value tasks; and (v) a reduction in instances of non-compliance or misuse of exemptions identified through improved verification precision. Such assessments should follow principles of phased implementation (Aarons et al., 2011; Damschroder et al., 2009). Furthermore, active collaboration with Google, within the parameters of existing government agreements, is advised to explore optimal configuration and potential fine-tuning of Gemini models for the specific nuances of Malaysian customs regulations, tariff nomenclature, and multilingual documentation (Gururangan et al., 2020; Howard & Ruder, 2018; Liao et al., 2023), which will be crucial for maximizing accuracy and relevance. Lastly, a significant investment in comprehensive training programs for officers is essential. This training should extend beyond basic AI tool usage to encompass effective interaction with Gemini-powered systems, critical interpretation of AI-generated suggestions, understanding AI limitations, and data ethics (Nonaka, 1994; OECD, various skills reports), as such programs are vital for fostering user acceptance and maximizing the "co-pilot" potential of the AI (Venkatesh et al., 2003).

For the broader Malaysian Government and policymakers, two key areas of focus are recommended. Firstly, ensuring that the necessary supporting digital infrastructure, robust data governance policies, and cybersecurity frameworks are firmly in place is paramount to facilitate the effective, secure, and ethical use of advanced AI like Gemini across public services. This includes addressing data residency, privacy, and cross-agency data sharing protocols (Puthucheary & Sivanathan, 2022; AlRyalat, Al-Hawari, & Al-Omari, 2021; NIST AI RMF), supported by national AI strategies (Ministry of Digital / MOSTI, Malaysia). Secondly, there should be a systematic evaluation of the outcomes, benefits, and challenges from the initial applications of Gemini within tax administration. These evidence-based insights should then inform broader strategies for deploying Gemini or similar LLMs across other public service sectors where complex text analysis, decision support, and enhanced efficiency are required (Walker, Damanpour, & Devece, 2010), employing a system thinking approach to anticipate wider impacts (Meadows, 2008).

### 7.4. Limitations of the Study

This study, while comprehensive in its review and synthesis, is subject to certain inherent considerations in its design and scope, as acknowledged in Chapter 3. Firstly, its findings are primarily based on a qualitative synthesis of existing literature and publicly available information, without the benefit of primary data collection from administrative officers or direct system prototyping. Secondly, the proposed framework is conceptual; its practical viability, precise performance metrics, and actual impact can only be determined through empirical testing and real-world implementation. Thirdly, the capabilities of Google's Gemini and other LLMs are rapidly evolving. While this study is based on knowledge available up to early 2025, future advancements may offer new possibilities or influence some of the assumptions made regarding specific functionalities. Finally, the study provides a high-level consideration of strategic aspects and ethical implications, each of which would warrant more in-depth investigation in subsequent phases of development and deployment.

### 7.5. Avenues for Future Research

The conceptual framework and findings presented in this study open up several avenues for future research that would be critical for advancing the practical application of Gemini in Malaysian tax administration and beyond. The most immediate need is for empirical research to validate the proposed conceptual framework. This would involve developing a prototype of the OCR/Gemini-powered system and conducting pilot studies within the tax authority to measure its impact on processing times, precision rates, officer productivity, and user satisfaction. Concurrently, future research should delve deeper into leveraging Gemini's multimodal capabilities, as detailed in Google AI publications and Bommasani et al. (2021) and noted for early 2025. This could involve investigating its utility in analyzing scanned documents that include images, diagrams (e.g., technical specifications of machinery), or even potentially connecting to visual inspection data of goods, further enriching the verification process. To ensure optimal selection and deployment, comparative studies evaluating the performance, cost-effectiveness, and suitability of Gemini against other leading LLMs or alternative AI techniques for the specific tasks involved in tariff verification within the Malaysian context, considering their respective states of development as of early 2025, would also be beneficial.

Beyond technical performance, it is crucial to undertake in-depth socio-technical studies examining the long-term impact of such AI systems on the workforce, including changes in job roles, skill requirements, organizational culture, and the human-AI collaboration dynamics within administrative bodies. Complementing this, research into the development of robust ethical auditing mechanisms specifically tailored for LLM-based systems in public administration is vital, focusing on ensuring fairness, accountability, transparency, and compliance with evolving AI regulations within the national context. Furthermore, investigating the integration of the Gemini-powered system with other existing or planned digital government platforms in the nation is essential to create a more seamless and data-driven public service ecosystem, potentially exploring synergies with technologies like blockchain for enhanced supply chain and trade document verification. Finally, comprehensive cost-benefit analyses for full-scale deployment are necessary. Future research into strategies for optimizing the scalability and computational efficiency of the system to manage costs effectively while expanding its application should particularly explore nuanced operational models such as 'tiered LLM usage.' This would involve investigating the feasibility of employing a spectrum of LLM capabilities potentially using less computationally intensive and therefore more resource-efficient LLM versions or models for initial, simpler tasks like basic data extraction confirmation from OCR outputs or routing straightforward queries, while reserving the full power of the advanced Gemini model (especially its renowned multimodal understanding and deep reasoning capabilities as understood in early 2025) for the more complex semantic analysis, regulatory interpretation, and resolution of considerations. Such a tiered approach could significantly optimize ongoing operational costs without compromising the effectiveness of critical verification stages, particularly when expanding its application.

## 8. Reference

Aarons, G. A., Hurlburt, M., & Horwitz, S. M. (2011). Advancing a conceptual model of evidence-based practice implementation in public service sectors. Administration and Policy in Mental Health and Mental Health Services Research, 38(1), 4-23.

Abdullah, N., & Zolkornain, Z. (2021). The Impact of Foreign Direct Investment on Economic Growth: The Case of Malaysia. Journal of Asian Finance, Economics and Business, 8(3), 1075-1083.

Acemoglu, D., & Restrepo, P. (2019). Automation and new tasks: How technology displaces and reinstates labor. Journal of Economic Perspectives, 33(2), 3-30.

Adadi, A., & Berrada, M. (2018). Peeking inside the black-box: A survey on explainable artificial intelligence (XAI). IEEE Access, 6, 52138-52160.

Aggarwal, C. C. (2018). Neural Networks and Deep Learning: A Textbook. Springer.

Al-Hashimi, M., & Kim, D. J. (2021). Factors affecting the adoption of artificial intelligence in the public sector: A systematic literature review. Government Information Quarterly, 38(4), 101613.

AlRyalat, M., Al-Hawari, M. A., & Al-Omari, M. K. (2021). Data governance in the era of big data: A systematic literature review. Journal of Big Data, 8(1), 1-31.

Alavi, M., & Leidner, D. E. (2001). Review: Knowledge management and knowledge management systems: Conceptual foundations and research issues. MIS Quarterly, 25(1), 107-136.

Alford, J., & Hughes, O. (2008). Public value pragmatism as the next phase of public management. The American Review of Public Administration, 38(2), 130-148.

Allen, B., & Gichoya, D. (2020). Building AI capacity in African governments: A framework for action. AI and Society, 35(4), 827-839.

Amershi, S., Weld, D., Vorvoreanu, M., Fourney, A., Nushi, B., Collisson, P., ... & Teevan, J. (2019). Guidelines for human-AI interaction. Proceedings of the 2019 CHI conference on human factors in computing systems, 1-13.

Ariely, D. (2008). Predictably Irrational: The Hidden Forces That Shape Our Decisions. HarperCollins.

Ariffin, A. S., & Ahmad, F. (2023). Regulating Artificial Intelligence in Malaysia: Legal and Ethical Challenges. Journal of Malaysian and Comparative Law, 50(2), 29-55.

Arun, S., & Heeks, R. (2019). Digital development: A research agenda. Information Technology for Development, 25(1), 1-19.

Athukorala, P. C. (2019). FDI and Development: The Case of Malaysia. Asian Development Review, 36(2), 93-119.

Autor, D. H. (2015). Why are there still so many jobs? The history and future of workplace automation. Journal of Economic Perspectives, 29(3), 3-30.

Barcevičius, E., Cibaitė, G., & Codagnone, C. (2019). Exploring the Evolving Normal of AI in Public Services: A Typology of AI Applications. Publications Office of the European Union.

Batura, O., & Lin, M. (2022). AI governance in ASEAN: Navigating between innovation, ethics, and geopolitics. The Pacific Review, 35(5), 789-816.

Benartzi, S., Beshears, J., Milkman, K. L., Sunstein, C. R., Thaler, R. H., Shankar, M., ... & Galing, S. (2017). Should governments invest more in nudging? Psychological Science, 28(8), 1041-1055.

Benbasat, I., & Lim, J. (2000). Information technology support for decision making: From "what-is" to "what-should-be". In the Blackwell handbook of principles of organizational behavior (pp. 421-438). Blackwell.

Bird, R. M. (2015). Tax Administration and Tax Reform: Reflections on Thirty Years of International Experience. In J. Alm & J. Martinez-Vazquez (Eds.), The Challenges of Tax Reform in a Globalized World (pp. 29-56). Emerald Group Publishing Limited.

Bommasani, R., Hudson, D. A., Adeli, E., Altman, R., Arora, S., von Arx, S., ... & Liang, P. (2021). On the opportunities and risks of foundation models. arXiv preprint arXiv:2108.07258.

Bostrom, R. P., & Heinen, J. S. (1977). MIS problems and failures: A socio-technical perspective. Part I: The causes. MIS Quarterly, 1(3), 17-32.

Brown, T. B., Mann, B., Ryder, N., Subbiah, M., Kaplan, J., Dhariwal, P., ... & Amodei, D. (2020). Language models are few-shot learners. Advances in neural information processing systems, 33, 1877-1901.

Brynjolfsson, E., & McAfee, A. (2017). Machine, Platform, Crowd: Harnessing Our Digital Future. W. W. Norton & Company.

Bryson, J. M., Crosby, B. C., & Bloomberg, L. (2014). Public value governance: A new paradigm for public management. Public Administration Review, 74(4), 445-452.

Buchanan, B. (2019). The Cybersecurity Dilemma: Hacking, Trust and Fear Between Nations. Oxford University Press.

Cabinet Office & Government Digital Service UK. (2023). A guide to using artificial intelligence in the public sector.

Chaisse, J. (2022). Artificial Intelligence and International Economic Law: A New Frontier for Theory and Practice. Hart Publishing.

Chalkidis, I., Fergadiotis, M., Malakasiotis, P., Aletras, N., & Androutsopoulos, I. (2020). LEGAL-BERT: The Muppets straight out of law school. arXiv preprint arXiv:2010.02559.

Choudrie, J., Islam, M. S., Wahid, F., & Bass, J. M. (2020). Information and communication technologies for development (ICT4D) research: A comprehensive literature review. Information Technology for Development, 26(3), 417-434.

Coiera, E. (2004). Four rules for the new IT. BMJ, 328(7449), 1147-1148.

Comiter, M. (2019). Attacking Artificial Intelligence: AI's Security Vulnerability and What Policymakers Can Do About It. Belfer Center for Science and International Affairs, Harvard Kennedy School.

Corteva, L. (2023). Artificial intelligence in customs: A game changer for trade facilitation and compliance? World Customs Journal, 17(1), 55-68.

Creswell, J. W., & Poth, C. N. (2016). Qualitative inquiry and research design: Choosing among five approaches. Sage publications.

Criado, J. I., & Gil-Garcia, J. R. (2019). Creating public value through smart technologies and strategies: From digital government to smart government and beyond. International Journal of Public Sector Management, 32(5), 438-450.

Damschroder, L. J., Aron, D. C., Keith, R. E., Kirsh, S. R., Alexander, J. A., & Lowery, J. C. (2009). Fostering implementation of health services research findings into practice: a consolidated framework for advancing implementation science. Implementation Science, 4(1), 50.

Dawes, S. S. (2009). Governance in the age of digital government: A research agenda. Journal of Public Administration Research and Theory, 19(1), 1-12.

De Witte, K., & Geys, B. (2011). Evaluating the efficiency of public service provision: A coherent and robust DEA-based methodology. European Journal of Operational Research, 209(1), 92-101.

Dignum, V. (2019). Responsible Artificial Intelligence: How to Develop and Use AI in a Responsible Way. Springer.

Dodge, J., Sap, M., Marasović, A., Agnew, W., Ilharco, G., Groeneveld, D., ... & Gardner, M. (2020). Documenting the state of the art in language modeling. arXiv preprint arXiv:2010.00102.

Doshi-Velez, F., & Kim, B. (2017). Towards a rigorous science of interpretable machine learning. arXiv preprint arXiv:1702.08608.

Dwivedi, Y. K., Hughes, L., Ismagilova, E., Aarts, G., Coombs, C., Crick, T., ... & Williams, M. D. (2021). Artificial Intelligence (AI): Multidisciplinary perspectives on emerging challenges, opportunities, and agenda for research, practice and policy. International Journal of Information Management, 57, 101994.

El-Haddadeh, R., Weerakkody, V., & Osmani, M. (2019). The impact of digital government transformation on citizen services: A systematic literature review. Information Polity, 24(2), 101-120.

Eloundou, T., Manning, S., Mishkin, P., & Rock, D. (2023). GPTs are GPTs: An Early Look at the Labor Market Impact Potential of Large Language Models. OpenAI.

Endsley, M. R. (2017). Toward a theory of situation awareness in dynamic systems. Human Factors, 37(1), 32-64.

Eom, S. B. (2009). Decision support systems research: current state and trends. International Journal of Information and Decision Sciences, 1(3), 213-235.

European Commission's High-Level Expert Group on AI. (2019). Ethics Guidelines for Trustworthy AI.

Fedi, L. (2018). Artificial intelligence and legal interpretation: a focus on customs tariff classification. Global Trade and Customs Journal, 13(7/8), 311-318.

Felbermayr, G., Teti, F., & Yalcin, E. (2021). On the complex relationship between tariff-classification uncertainty and international trade. European Economic Review, 131, 103602.

Floridi, L. (2019). Establishing the rules for building trustworthy AI. Nature Machine Intelligence, 1(6), 261-262.

Floridi, L., Cowls, J., Beltramini, M., Saunders, D., & Vayena, E. (2018). An ethical framework for a good AI society: opportunities, risks, principles, and recommendations. AI and Society, 33(4), 689-707.

Flyverbom, M., Deibert, R., & Matzner, T. (2019). The governance of digital technology: The contested Rrglobal landscape. Global Policy, 10(4), 529-540.

Frey, C. B., & Osborne, M. A. (2017). The future of employment: How susceptible are jobs to computerization? Technological Forecasting and Social Change, 114, 254-280.

Fullan, M. (2006). Change theory: A force for school improvement. Centre for Strategic Education.

Grudin, J. (2009). AI and HCI: Two fields divided by a common focus. IEEE Intelligent systems, 24(5), 48-57.

Guidotti, R., Monreale, A., Ruggieri, S., Turini, F., Giannotti, F., & Pedreschi, D. (2018). A survey of methods for explaining black box models. ACM Computing Surveys (CSUR), 51(5), 1-42.

Gururangan, S., Marasović, A., Swayamdipta, S., Lo, K., Beltagy, I., Downey, D., & Smith, N. A. (2020). Don't stop pretraining: Adapt language models to domains and tasks. Proceedings of the 58th Annual Meeting of the Association for Computational Linguistics, 8342-8360.

Hartley, J. (2013). The creation and capture of public value: A new research agenda for public management. International Journal of Public Sector Management, 26(1), 61-72.

Heeks, R. (2006). Implementing and managing eGovernment: An international text. Sage Publications.

Helbing, D. (2019). Societal, economic, ethical and legal challenges of the digital revolution: From big data to deep learning, artificial intelligence, and manipulative technologies. In Towards digital enlightenment (pp. 47-72). Springer, Cham.

Hendrycks, D., Burns, C., Kadavath, S., Arora, A., Basart, S., Tang, E., ... & Steinhardt, J. (2021). Measuring massive multitask language understanding. Proceedings of the International Conference on Learning Representations (ICLR).

Hollnagel, E., & Woods, D. D. (2005). Joint Cognitive Systems: Foundations of Cognitive Systems Engineering. CRC Press.

Holzinger, A., Carrington, A., & Müller, H. (2020). Measuring the quality of explanations: The system causability scale (SCS). KI-Künstliche Intelligenz, 34, 193-198.

Howard, J., & Ruder, S. (2018). Universal language model fine-tuning for text classification. Proceedings of the 56th Annual Meeting of the Association for Computational Linguistics (Volume 1: Long Papers), 328-339.

Janssen, M., van der Voort, H., & Wahyudi, A. (2017). Factors influencing data governance in the public sector. Information Polity, 22(2-3), 143-158.

Jobin, A., Ienca, M., & Vayena, E. (2019). The global landscape of AI ethics guidelines. Nature Machine Intelligence, 1(9), 389-399.

Katzenbach, C., & Ulbricht, L. (2019). Algorithmic governance. Internet Policy Review, 8(4), 1-18.

Keen, M., & Slemrod, J. (2017). Optimal Tax Administration. Journal of Public Economics, 152, 133-142.

Klanrit, P., Lee, W. Y., & Kim, H. (2023). A Survey on Deep Learning for Legal Document Analytics. ACM Computing Surveys, 55(8), 1-38.

Klein, G. (2008). Naturalistic decision making. Human Factors, 50(3), 456-460.

Klievink, B., Veeneman, W. W. (2018). Creating value with big data analytics in public administration: A framework for aligning purposes, capabilities, and context. Government Information Quarterly, 35(3), 470-479.

Koh, C. E., Ryan, S., & Prybutok, V. R. (2005). Creating value through managing knowledge in an e-government to constituency (G2C) environment. Journal of Computer Information Systems, 45(4), 32-41.

Kuziemsky, C. E., & Varpio, L. (2022). The Use of Artificial Intelligence in Public Administration: A Systematic Literature Review. Government Information Quarterly, 39(3), 101693.

Lewis, M., Liu, Y., Goyal, N., Ghazvininejad, M., Mohamed, A., Levy, O., ... & Zettlemoyer, L. (2020). BART: Denoising sequence-to-sequence pre-training for natural language generation, translation, and comprehension. arXiv preprint arXiv:1910.13461.

Liang, P., Bommasani, R., Lee, T., Tsipras, D., Soylu, D., Yasunaga, M., ... & Koreeda, Y. (2022). Report on the 2021 Workshop on Large Language Models: Invited Talks and Discussion Panels. Stanford University Human-Centered Artificial Intelligence (HAI).

Liao, G., He, Q., Gao, M., Li, Z., & Chua, T. S. (2023). Specialized Large Language Models: A Survey. arXiv preprint arXiv:2310.09188.

Malaysian Administrative Modernisation and Management Planning Unit (MAMPU). (2021). Pelan Strategik Pendigitalan Sektor Awam 2021-2025 (Public Sector Digitalisation Strategic Plan 2021-2025).

Meadows, D. H. (2008). Thinking in Systems: A Primer. Chelsea Green Publishing.

Medaglia, R., & Zheng, L. (2017). Public value creation by ICT in public sector: A research and practice agenda. Government Information Quarterly, 34(2), 167-171.

Mehr, H. (2017). Artificial Intelligence for Citizen Services and Government. Ash Center for Democratic Governance and Innovation, Harvard Kennedy School.

Miles, M. B., Huberman, A. M., & Saldaña, J. (2018). Qualitative data analysis: A methods sourcebook. Sage publications.

Miller, T. (2019). Explanation in artificial intelligence: Insights from the social sciences. Artificial Intelligence, 267, 1-38.

Ministry of Investment, Trade and Industry (MITI). (2023). New Industrial Master Plan 2030 (NIMP 2030). MITI.

Moore, M. H. (1995). Creating Public Value: Strategic Management in Government. Harvard University Press.

Mori, S., Suen, C. Y., & Yamamoto, K. (1992). Historical review of OCR research and development. Proceedings of the IEEE, 80(7), 1029-1058.

Mumford, E. (2006). The story of socio-technical design: reflections on its successes, failures and potential. Information Systems Journal, 16(4), 317-342.

Ndung'u, B., & Waema, T. M. (2019). Digital Transformation of Public Services in Africa. Routledge.

Nonaka, I. (1994). A dynamic theory of organizational knowledge creation. Organization Science, 5(1), 14-37.

O'Dell, C., & Hubert, C. (2011). The New Edge in Knowledge: How Knowledge Management is Changing the Way We Do Business. John Wiley & Sons.

OECD. (2019). Artificial Intelligence in Society. OECD Publishing.

OpenAI. (2023). GPT-4 Technical Report. https://arxiv.org/abs/2303.08774

Parasuraman, R., Sheridan, T. B., & Wickens, C. D. (2000). A model for types and levels of human interaction with automation. IEEE Transactions on systems, man, and cybernetics-Part A: Systems and Humans, 30(3), 286-297.

Pauwelyn, J. (2020). The End of IIAs as We Know Them? How AI and Big Data are (Quietly) Reshaping International Investment Law and Arbitration. Journal of International Economic Law, 23(3), 557-585.

Piccinini, E., Hanelt, A., Gregory, R. W., & Kolbe, L. M. (2015). Transforming industrial business: the impact of digital transformation on automotive manufacturers. In Proceedings of the International Conference on Information Systems (ICIS), Article 1.

Pohle, J., & Thiel, T. (2020). Digital sovereignty. Internet Policy Review, 9(4), 1-11.

Polanyi, M. (1966). The Tacit Dimension. University of Chicago Press.

Puthucheary, S. D., & Sivanathan, S. S. (2022). Data Governance in Malaysia: An Overview of the Legal Framework. Asian Journal of Law and Governance, 4(1), 1-14.

Rai, A. (2020). Explainable AI: from black box to glass box. Journal of the Academy of Marketing Science, 48(1), 137-141.

Rose, J., Persson, J. S., & Heeager, L. T. (2015). How ideal-types of IS projects impact project management. Project Management Journal, 46(5), 46-59.

Rowland, C., & Podnar, I. (Eds.). (2021). Cybersecurity for Artificial Intelligence. Springer.

Rudin, C. (2019). Stop explaining black box machine learning models for high stakes decisions and use interpretable models instead. Nature Machine Intelligence, 1(5), 206-215.

Russell, S. J., & Norvig, P. (2020). Artificial Intelligence: A Modern Approach (4th ed.). Pearson.

Sawyer, A. J. (Ed.). (2018). Taxation in ASEAN and China: A Guide for the Tax Practitioner and Business Executive (5th ed.). CCH Asia Pte Ltd.

Shneiderman, B. (2020). Human-centered AI: Reliable, safe & trustworthy. International Journal of Human–Computer Interaction, 36(6), 495-504.

Smith, R. (2007). An overview of the Tesseract OCR engine. Proceedings of the Ninth International Conference on Document Analysis and Recognition (ICDAR 2007) (Vol. 2, pp. 629-633). IEEE.

Souza, L. A., Batista, F., & Silva, T. P. (2021). Intelligent document processing: A systematic literature review. Expert Systems with Applications, 183, 115335.

Sterman, J. D. (2000). Business Dynamics: Systems Thinking and Modeling for a Complex World. Irwin/McGraw-Hill.

Sunstein, C. R. (2016). The Ethics of Nudging. Yale University Press.

Susser, D., Roessler, B., & Nissenbaum, H. (2019). Online manipulation: Hidden influences in a digital world. Georgetown Law Technology Review, 4(1), 1-45.

Taleb, N. N. (2012). Antifragile: Things That Gain from Disorder. Random House.

Thaler, R. H., & Sunstein, C. R. (2008). Nudge: Improving Decisions About Health, Wealth, and Happiness. Yale University Press.

Twizeyimana, J. D., & Andersson, A. (2019). The public value of E-Government – A literature review. Government Information Quarterly, 36(2), 167-178.

Veale, M., & Borgesius, F. Z. (2021). Demystifying the Draft EU Artificial Intelligence Act. Computer Law Review International, 22(4), 97-112.

Venkatesh, V., Morris, M. G., Davis, G. B., & Davis, F. D. (2003). User acceptance of information technology: Toward a unified view. MIS Quarterly, 27(3), 425-478.

Wagner, B. (2018). Digital sovereignty: The new frontier for state power. Carr Center for Human Rights Policy Discussion Paper Series. Harvard Kennedy School.

Walker, R. M., Damanpour, F., & Devece, C. A. (2010). Management innovation and organizational performance: The mediating effect of performance management. Journal of Public Administration Research and Theory, 21(2), 367-386.

Wang, D., Govindarajulu, U. S., Kankanhalli, A., & Siau, K. L. (2021). Explainable AI in Human-AI Decision-Making: A Moderated Mediation Model of Explanation Type, System Trust, and Task Complexity. Information Systems Research, 32(4), 1221-1242.

Waugh, L. R. (1997). The linguistic structure of the Harmonized System. Terminology. International Journal of Theoretical and Applied Issues in Specialized Communication, 4(1), 47-68.

Wolf, T., Debut, L., Sanh, V., Chaumond, J., Delangue, C., Moi, A., ... & Rush, A. M. (2020). Transformers: State-of-the-art natural language processing. Proceedings of the 2020 Conference on Empirical Methods in Natural Language Processing: System Demonstrations, 38-45.

Yin, R. K. (2017). Case study research and applications: Design and methods. Sage publications.

Zhao, W. X., Zhou, K., Li, J., Tang, T., Wang, X., Hou, Y., ... & Wen, J. R. (2023). A survey of large language models. arXiv preprint arXiv:2303.18223.

Zhong, H., Cui, L., Zhang, Y., Wang, C., & Huang, S. (2020). How does NLP benefit legal system: A summary of legal artificial intelligence. arXiv preprint arXiv:2004.07031.

van den Hoven, J., Vermaas, P. E., & van de Poel, I. (Eds.). (2015). Handbook of ethics, values, and technological design. Springer.

## 9. Appendix

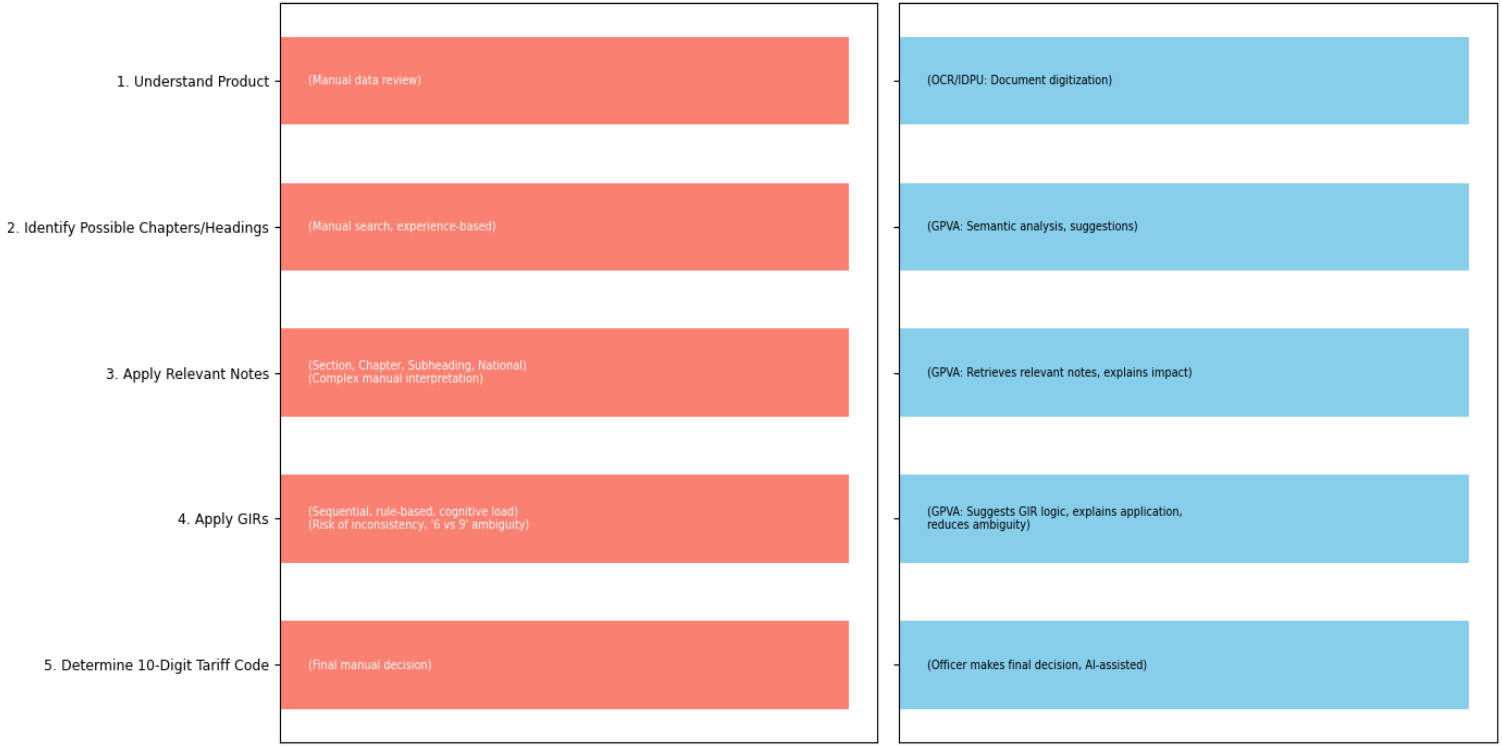

Figure 14: Differentiation between tariff classification workflow

Figure 15: Visual simulation of GPVA Classification Assistance